\documentclass[11pt]{article}
\usepackage{amsmath,amssymb,amsfonts} 
\usepackage{graphicx,times}
\usepackage{natbib}
\usepackage{mathrsfs}
\usepackage{color}
\providecommand{\bm}[1]{\mbox{\boldmath$#1$\unboldmath}}

\begin{document}

\title{A Computational Study of Particle Deposition Patterns
from a Circular Laminar Jet}

\author{James Q.~Feng}

\maketitle

\begin{center}
{Optomec, Inc., 2575 University Avenue, \#135, St. Paul, Minnesota 55114, USA} \\
{{\bf email:} jfeng@optomec.com}
\end{center}

\vspace{10 mm}

\noindent\rule{160 mm}{0.5 pt}

\begin{center}
{\bf Abstract}
\end{center}

Particle deposition patterns on the plate of inertial impactor 
with circular laminar jet
are investigated numerically 
with a Lagrangian solver implemented within the 
framework of the 
OpenFOAM$^{\circledR}$ 
CFD package.
Effects of taper angle of the nozzle channel 
and jet-to-plate distance are evaluated. 
The results show that tapered nozzle tends to deposit 
more particles toward the circular spot edge than straight nozzle.
At jet Reynolds number $Re = 1132$,
a tapered nozzle 
deposits particles to form a pattern with a high density ring toward 
the deposition spot edge, especially for particle Stokes number $St > St_{50}$, 
which is absent with a straight nozzle.
Increasing the jet-to-plate distance tends to reduce the value of 
particle density peak near deposition spot edge. 
Reducing $Re$ to $283$ 
(e.g., for $300$ ccm flow through a $1.5$ mm diameter jet nozzle) 
yields particle deposition patterns without the high density ring 
at the deposition spot edge 
when the same tapered nozzle is used.
The particle deposition patterns with the straight nozzle at $Re = 283$
exhibit further reduced particle density around the spot edge such that
the particle density profile appears more or less like a Gaussian function.
In general, the effect of reducing $Re$ on particle deposition pattern
seems to be similar to increasing the 
jet-to-plate distance. 
The computed particle deposition efficiency $\eta$ shows 
the fact that very fine particles with extremely small 
values of $St$ near the jet axis always impact the center of plate,
indicating that the value of $\eta$ does not approach zero 
with a substantial reduction of $St$.
Such a ``small particle contamination'' typically amounts to $\sim 10\%$ of 
small particles (with $\sqrt{St} < 0.1$)
at $Re \sim 1000$ and $\sim 5\%$ at $Re \sim 300$, 
which may not be negligible in data analysis with 
inertial impactor measurement. 

\vspace{2 mm}

\noindent\rule{160 mm}{0.5 pt}

\vspace{8 mm}

%\noindent
%{{\bf Subject:} ...} \\
%{{\bf Keywords:} ...}

\section{Introduction}
Many authors have studied particle impaction behavior with circular jets
mainly for its application in aerosol particle classification 
by aerodynamic size with the cascade impactors 
\citep[e.g.,][]{andersen1966, marple1976, hering1995}.
The original inertial impaction theory was 
presented by \cite{ranz1952}.
Practical designs of the inertial impactors have been guided by
numerically computing the flow field governed by the Navier-Stokes equations
and then integrating equations governing the particle motions for  
analysis of particle trajectories 
\citep[cf. ][]{marple1970, marple1974, huang2001}, 
Based on a thorough parametric study, 
\cite{marple1970, marple1974, marple1975}
found that sharp cutoff deposition efficiency curves 
can be obtained when the jet Reynolds number is between $500$ and $3000$.
The gravity effect on collection efficiency of large particles in 
the low-velocity inertial impactor
was demonstrated experimentally by \cite{may1975}.
Both numerical and experimental studies of the gravity effect on 
particle collection efficiency in inertial impactors were
carried out by \cite{huang2001}.
However, the study of particle deposition patterns with 
a circular laminar jet could only be found in a publication by
\cite{sethi1993} with laboratory experiments for one geometric configuration.

With the Aerosol Jet$^{\circledR}$ direct-write technology,
functional ink is aerosolized via an atomizer and transported 
as a dense mist of microdroplets (usually about $50$ nL/cc), 
wrapped with a sheath gas through
a nozzle with an appropriate orifice, and then deposited 
onto the substrate 
by the mechanism of inertial impaction with an impinging jet flow
\citep[cf. ][]{renn2006, renn2007, 
renn2009}, 
enabling precision high-aspect-ratio material deposition for 
a variety of scientific and industrial applications 
\citep[cf. ][]{hedges2007, 
kahn2007, renn2010, 
christenson2011, paulsen2012}.
For well-controlled high-precision material deposition,
the aerosol mist flow impacting onto the substrate is
maintained in the steady laminar regime for
Aerosol Jet$^{\circledR}$
printing.
An in-depth understanding 
of deposition patterns
of microdroplets (typically with diameters of a few microns)
with a circular laminar jet
is important for
Aerosol Jet$^{\circledR}$
deposition nozzle design as well as process development. 

In the present work, a method for evaluating 
particle deposition patterns is developed with 
computational analysis using a Lagrangian solver implemented within  
the framework of OpenFOAM$^{\circledR}$ CFD package
(www.openfoam.com/documentation/user-guide/).
In what follows, the computational methodology  
is presented in section 2, 
and then results and discussion in section 3 
for cases of straight nozzle ($\phi = 0$) and tapered nozzle ($\phi = 15^o$).
Finally, the conclusions are summarized in section 4.

\begin{figure}[t!] \label{schematic}
\includegraphics[scale=0.6]{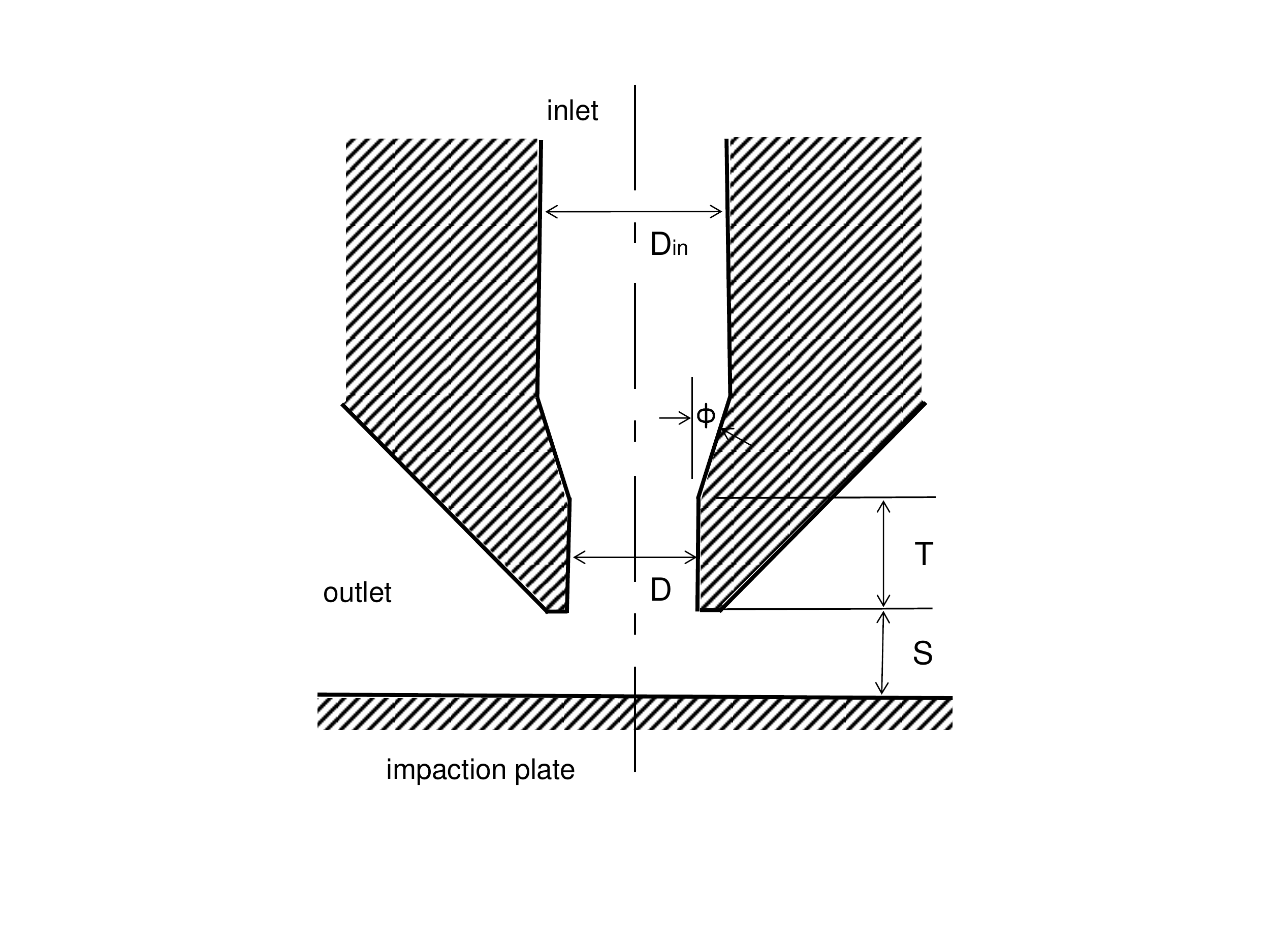}
\caption{Schematic of 
the circular-jet impactor geometric configuration.}
\end{figure}

\section{Computational Methodology}
As schematically shown in Fig. 1, the inertial impactor consists of 
a nozzle with an circular orifice of diameter $D$ and an impaction plate
of much larger diameter located at a ``jet-to-plate'' distance $S$.
Moreover, geometric parameters such as nozzle throat length $T$, 
half angle of the tapering section $\phi$, etc. can also influence
the particle impaction behavior.
With a given geometric configuration,
the particle-laden mist flow may be generally assumed as incompressible 
(with the flow velocity much less than the speed of sound), 
laminar (with jet Reynolds number less than $1500$), and steady 
(with the jet-to-plate distance $S$ comparable to $D$).
Thus, the flow field is governed by the continuity equation for
incompressible flow
\begin{equation} \label{continuity}
\nabla \cdot \bm{u} = 0 
\, \, ,
\end{equation}
and steady flow momentum equation (also known as the 
Narvier-Stokes equations for steady flow)
\begin{equation} \label{momentum}
\nabla \cdot (\bm{u} \bm{u}) + \nabla \cdot (\nu \nabla \bm{u}) = - \nabla p 
\, \, ,
\end{equation}
where $\bm{u}$ is the vector field of flow velocity and 
$p$ the kinematic pressure (which comes from the thermodynamic pressure
divided by the constant fluid density of the carrier gas $\rho$)
with $\nu$ denoting the (constant) kinematic viscosity
of the carrier gas (assuming the mist of particles is not too dense 
to cause fluid viscosity variation).
The solution of flow field governed by (\ref{continuity}) and
(\ref{momentum}) can be computed with the 
``simpleFoam'' solver (implemented for steady incompressible flow
using the SIMPLE algorithm) in 
the OpenFOAM$^{\circledR}$ CFD package.
If length is measured in units of the nozzle orifice diameter $D$,
velocity $\bm{u}$ in units of $U \equiv 4 Q/(\pi \, D^2)$ 
with $Q$ denoting the 
volumetric flow rate entering the impactor, 
and kinematic pressure $p$ in units of $U^2$,
the nondimensionalized (\ref{momentum}) would have $\nu$ 
replaced by $1/Re$ where the jet Reynolds number $Re$ is 
defined as
\begin{equation} \label{reynolds}
Re \equiv \frac{U \, D}{\nu}
= \frac{\rho \, U \, D}{\mu}
\, \, ,
\end{equation}
with $\mu$ denoting the dynamic viscosity of the carrier gas. 
For the present problem (cf. Fig. 1), there are three types of boundaries: 
inlet, outlet, and walls.  
The boundary conditions at inlet are simply
``zeroGradient'' type for $p$ (namely, $\bm{n} \cdot \nabla p = 0$)
and ``flowRateInletVelocity'' type with a specified ``volumetricFlowRate'' 
$Q$ for $\bm{u}$, which is equivalent to having a plug flow at the inlet
as if the flow is coming from a large upstream open volume.
At outlet, 
``fixedValue'' type for $p$ ($= 0$) and 
``zeroGradient'' type for $\bm{u}$ are applied (namely, 
$\bm{n}\bm{n} : \nabla \bm{u} = 0$);
and at walls,  
``zeroGradient'' type for $p$ and
``fixedValue'' type for $\bm{u}$ ($= \bm{0}$) are used.

Once the flow velocity field $\bm{u}$ is computed, 
the position vector of each particle $\bm{x_p}$ in a Lagrangian frame
can be calculated from the equations of motion
\begin{equation} \label{motion_x}
\frac{d \bm{x_p}}{dt} =  \bm{u_p}
\, \, ,
\end{equation}
and
\begin{equation} \label{motion_u}
m_p \frac{d \bm{u_p}}{dt} = \sum \bm{f_p} 
\, \, ,
\end{equation}
where $\bm{u_p}$ denotes the particle velocity at 
position $\bm{x_p}$ and time $t$,
$m_p$ the particle mass,
and $\bm{f_p}$ the forces acting on the particle.
The OpenFOAM$^{\circledR}$ CFD package 
contains a ``basicKinematicCloud'' class to introduce kinematic 
parcels and to track the parcel positions according to the specified forces 
$\sum \bm{f_p}$.
In general, a parcel is a computational particle, which may contain multiple
actual particles depending upon the model specifications, to  
reduce the computational burden for tracking large number of individual
particles. When the number of particles is not large, a parcel can be 
computed as an individual actual particle (as in the present work). 

For 
Aerosol Jet$^{\circledR}$
printing, the mist of ink droplets usually contains ink 
of about $50$ nL/cc, or the ink volume fraction of $5 \times 10^{-5}$.
Thus, the droplets can be considered far enough apart that 
each droplet behave as an isolated spherical particle in the mist flow.
Because the ink droplets suitable for  
Aerosol Jet$^{\circledR}$ 
printing typically have diameters in the range of 1 to 5 $\mu$m,
Browniian diffusion effect should be negligible.
Thus, the dominant forces acting on each particle are the 
drag $\bm{f_d}$ due to the relative motion in fluid and 
the gravitational force $m_p \bm{g}$,
i.e., $\sum \bm{f_p} = \bm{f_d} + m_p \bm{g}$.
In the present work, the particles are assumed to be spheres to 
represent ink droplets; thus, ``sphereDrag'' and 
``gravity'' are specified as the 
``particleForces'' in the ``kinematicCloudProperties''.
In OpenFOAM-2.4.0 (which is used in the present work),
the ``sphereDrag'' $\bm{f_d}$ is computed according to
\begin{equation} \label{f_d}
\bm{f_d} = \frac{3\, m_p\, \mu\,  C_d \, Re_p}{4 \rho_p\, d_p^2} 
(\bm{u} - \bm{u_p}) + \frac{\pi \rho_p d_p^3}{6} \bm{g}
\, \, ,
\end{equation}
where $\rho_p$ denotes the 
particle density, $d_p$ the particle diameter, and $\bm{g}$ 
($= 9.81$ m s$^{-2}$) 
the gravitational acceleration in the axial flow velocity direction.
The drag coefficient $C_d$ is calculated according to
\begin{equation} \label{C_d1}
C_d = \left\{ \, 
\begin{split}
  & \frac{24}{Re_p} \left(1 + \frac{1}{6} Re_p^{2/3} \right)
\, \, \mathrm{ for } \, \, Re_p \le 1000 
\\
\\
  & 0.424
\, \, \mathrm{ for } \, \, Re_p > 1000 
\end{split}
\right.
\, \, \quad ,
\end{equation}
with the particle Reynolds number defined as
\begin{equation} \label{Re_p}
Re_p \equiv \frac{\rho\, d_p \, |\bm{u_p} - \bm{u}|}{\mu}
= \frac{d_p \, |\bm{u_p} - \bm{u}|}{\nu}
\, \,  .
\end{equation}

To keep the model theoretically clean, a ``Lagrangian'' solver is 
implemented within the
OpenFOAM$^{\circledR}$ 
framework such that the presence of kinematic cloud parcels does not 
disturb the given steady flow field obtained from the ``simpleFoam'' 
computation,
while the motion of parcels is determined by 
solving (\ref{motion_x})--(\ref{f_d}) from the given 
flow field $\bm{u}$.
It should be noted that (\ref{f_d}) in the 
OpenFOAM$^{\circledR}$ 
implementation
does not contain
the Cunningham slip correction factor $C_c$ in the denominator as usually seen
in the aerosol science literature. 
Therefore, it is added by modifying 
the source code of ``SphereDragForce.C'' according to 
\begin{equation} \label{C_c}
C_c = 1 + \frac{\lambda}{d_p} \left[A_1 + A_2 
\exp\left(\frac{-A_3 \, d_p}{\lambda} \right) \right]
\, \,  .
\end{equation}
where $\lambda$ is the mean free path of the gas 
(which is about $0.065$ $\mu$m at $25$ $^o$C) with 
$A_1 = 2.514$, $A_2 = 0.8$, and $A_3 = 0.55$ \citep{friedlander1977}.

With the
OpenFOAM$^{\circledR}$ 
Lagrangian solver,
particles can be introduced in flow by several built-in ``injectionModels'',
among which the type of ``manualInjection'' allows 
particles (one per parcel) of specified diameter to be injected at 
specified positions inside problem domain. 
In the present work, a set of particles of identical properties 
is placed near the flow inlet at known radial positions 
from the axis of symmetry 
with given spacing,
e.g., at $\hat{r}_i = i \times \Delta \hat{r}$ with $i = 0, 1, 2, ...$ ,
with the particle initial velocity specified to match that of the inlet
plug flow.
The radial positions of individual particles $r_i$ can be determined 
from the ``patchPostProcessing'' data file containing 
those particle deposition positions on the impaction plate.
Then, assuming the particle concentration is uniform at the nozzle inlet,
a dimensionless surface particle density 
$\sigma$ at a given radial position $r_i$ on the 
impaction plate can be calculated as 
\begin{equation} \label{p-density}
\sigma_i = 
\frac{\hat{r}_{i+1}^2-\hat{r}_{i-1}^2}{r_{i+1}^2-r_{i-1}^2}
\, \,  \mathrm{ for } \, i \ne 0 \, \, 
\mathrm{ and } \, \, \sigma_0 = 2 \sigma_1 - \sigma_2 \, \, .
\end{equation}
Noteworthy here is that the density of deposited particles on
the impaction plate $\sigma$ given by (\ref{p-density}) 
is evaluated as inversely proportional to
the change of relative spacing between 
neighboring particles,
with $\hat{r}_i$ and $r_i$ denoting the beginning 
(at the inlet) and ending (at the impaction plate) positions of 
the trajectory of particle $i$.
The number of particles arriving the impaction plate 
within the ring defined by $r_{i-1}$ and $r_{i+1}$ 
is expected to be conserved, i.e., being the same as 
that at inlet with plug flow within the ring defined by 
$\hat{r}_{i-1}$ and $\hat{r}_{i+1}$. 

For a given $d_p$, there is a critical radius $\hat{r}_c$ 
($\le D_{in} / 2$) at nozzle inlet beyond which
the particles would not deposit onto the impaction plate; they
exit through the outlet boundary.
If the particle concentration and flow velocity profile are 
assumed to be uniform at inlet (as consistent with the specified plug flow 
boundary condition),
the deposition efficiency can be determined as 
\begin{equation} \label{eta}
\eta = \left(\frac{2 \, \hat{r}_c}{D_{in}}\right)^2 
\, \,  ,
\end{equation}
where $D_{in}$ is the diameter of nozzle inlet (cf. Fig. 1).

The value of the Stokes number, defined as the ratio of 
the particle stopping distance and the radius of the 
nozzle orifice ($D/2$), is written as \citep{fuchs1964}
\begin{equation} \label{stk}
St = \frac{\rho_p \, U \, C_c \, d_p^2}{9 \, \mu \, D}
\, \,  .
\end{equation}
In the literature of inertial impactors, 
curves are usually presented in terms the particle deposition 
efficiency $\eta$ versus $\sqrt{St}$, where $\sqrt{St}$ is considered as 
the dimensionless particle diameter.   

With the OpenFOAM$^{\circledR}$ ``kinematicCloud'',
several interaction models between parcel and boundary patch are 
available.
Although particle rebound can be modeled with appropriately specified
elasticity and restitution coefficient, 
it is much simpler to just assume the particle remains where it 
contacts the surface. 
In the present work, 
the mode of local interactions between particles and boundaries, such as 
impaction plate, walls, and outlet, is simply specified as ``stick''
in the  
OpenFOAM$^{\circledR}$ 
input file,
which is especially reasonable for the
Aerosol Jet$^{\circledR}$ 
printing, 
where the particles are actually the liquid microdroplets of ink materials.
Thus, droplet rebounding and splashing are not considered in 
the computation, for simplicity.

\section{Results and Discussion}
For convenience of comparison, the nominal settings 
in the geometric configuration shown in Fig. 1 (similar to
that used by \cite{sethi1993}) 
are $D = 1.5$ mm, $T/D = 2.0$, with various $S/D$ and $\phi$.
The jet Reynolds number $Re$ can be varied by changing the 
volumetric flow rate $Q$ at inlet.
For the carrier gas, the value of $\mu$ is taken as 
$1.8 \times 10^{-5}$  N s m$^{-2}$
and $\rho$ as
$1.2$ kg m$^{-3}$, as typical values for nitrogen 
under ambient temperature and pressure. 
Thus, we have $Re = 1132$ for a flow rate of $Q = 1200$ ccm
with $U \equiv 4 Q/(\pi D^2)$ $= 11.3$ m s$^{-1}$.
The value of particle density $\rho_p$ is assumed to be $10^3$ kg m$^{-3}$;
therefore, the spherical particle diameter $d_p$ is the same as
the ``aerodynamic diameter''.

\begin{figure}[t!] \label{streamline}
\includegraphics[scale=0.56]{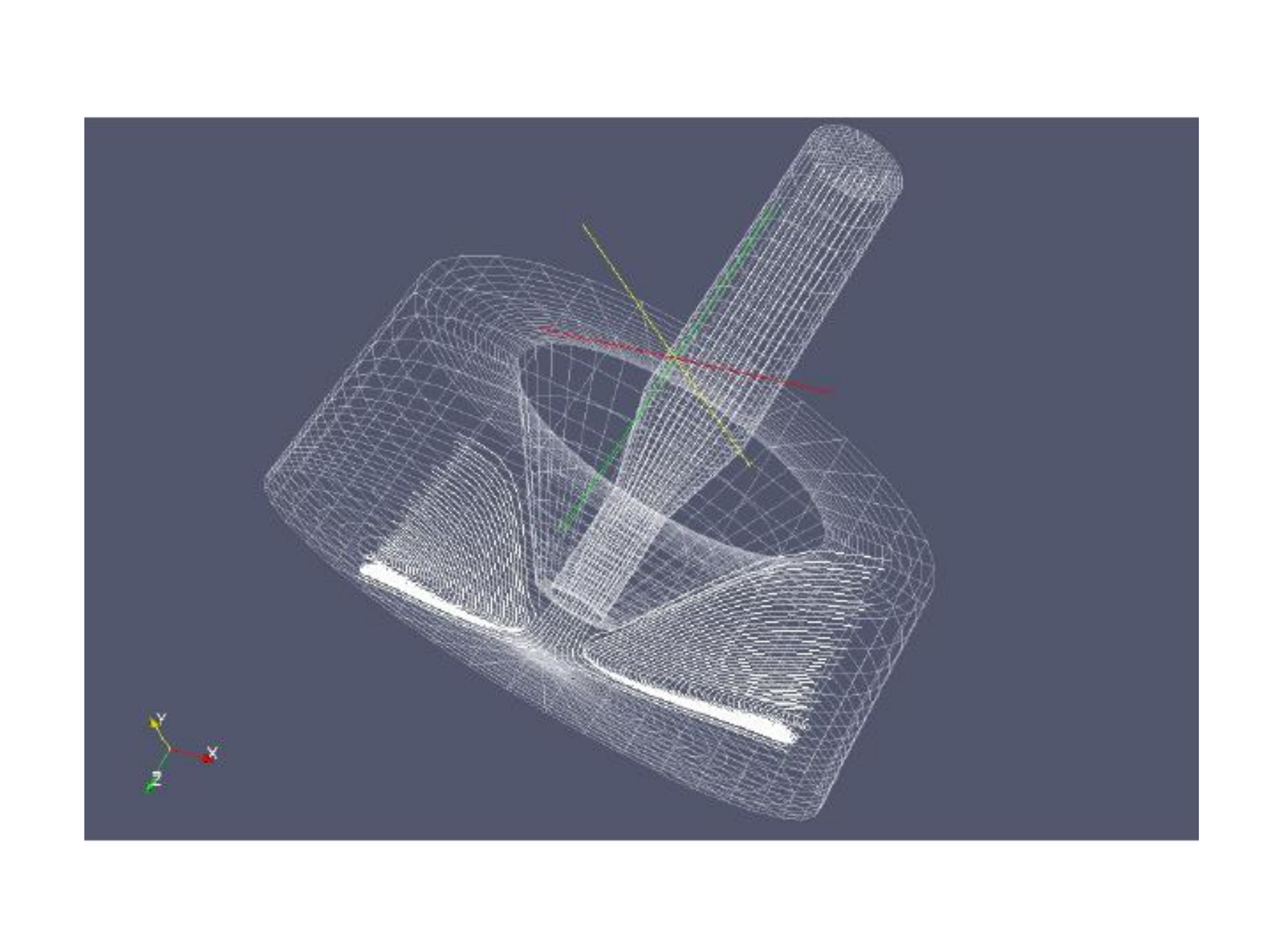}
\caption{Streamlines of flow in a circular-jet impactor (on the xz-plane)
for $Re = 1132$ with $D = 1.5$ mm and $D_{in} = 3$ mm, 
$S/D = 1$, $T/D = 2$, and $\phi = 15^o$. 
The three-dimensional mesh of problem domain generated with ``blockMesh'' 
is also shown in the plot. The outlet boundary is at radius 
equal to $5 \times D$.}
\end{figure}

Full three-dimensional mesh is used in the present problem such that
particles can be placed with adequate spacing in between,
to avoid particle-particle interactions 
in the ``basicKinematicCloud'' class of 
OpenFOAM$^{\circledR}$ package. 
Fig. 2 shows the streamline plot of ``simpleFoam'' computational result 
for the geometric configuration of $S/D = 1$, $T/D = 2$, and $\phi = 15^o$,
along with the three-dimensional mesh used for the computation.
The ``blockMesh'' utility of 
OpenFOAM$^{\circledR}$
is used for generating the high-quality hexahedral mesh 
with finite-volume cells 
in the nozzle and impaction regions being around $70$ $\mu$m,
which is determined to be adequate for accurately resolving
the laminar flow field of impinging jet while being much larger than the 
particle diameter as desired for Lagrangian tracking of discrete particles. 
The steady jet impinging flow structure appears fairly similar to that
shown for $Re = 1000$
in an independent study with a different computational methodology 
by \cite{feng2015}. 

\begin{table}
\caption{The values of 
$St$, $\sqrt{St}$, 
$\tau$ (ms), and 
$L_s$ (mm)
for various particle diameter $d_p$ at
$Re = 1132$ with $D = 1.5$ mm.}
\begin{center}
\begin{tabular*}{0.75\textwidth}{@{\extracolsep{\fill}} lcccc}
\hline
\hline
\\ %\vspace{ 1 mm }
  $d_p$ ($\mu$m) & $St$ & $\sqrt{St}$ & $\tau$ (ms) & $L_s$ (mm)  \\
\hline
\\
  0.5 & $0.0155$ & $0.1244$ & $1.02 \times 10^{-3}$ & $0.0116$  \\
  1 & $0.0542$ & $0.2328$ & $3.59 \times 10^{-3}$ & $0.0406$  \\
  1.5 & $0.1162$ & $0.3409$ & $7.70 \times 10^{-3}$ &  $0.0872$  \\
  1.75 & $0.1560$ & $0.3949$ & $0.0103$ & $0.1170$  \\
  2 & $0.2015$ & $0.4489$ & $0.0134$ & $0.1511$  \\
  3 & $0.4420$ & $0.6648$ & $0.0293$ & $0.3315$  \\
  4 & $0.7756$ & $0.8807$ & $0.0514$ & $0.5817$  \\
  5 & $1.2024$ & $1.0966$ & $0.0797$ & $0.9018$ \\ 
\hline
\hline
\end{tabular*}
\end{center}
\end{table}

For convenience, Table 1 illustrates the values of 
the Stokes number $St$, the dimensionless particle size $\sqrt{St}$,
along with the often referred to particle motion parameters 
\citep[e.g.,][]{fuchs1964} such as 
characteristic time (or relaxation time) of particles 
in response to nonuniform rectilinear flow
$\tau \equiv St \, D/(2 \, U)$ and 
stop distance $L_s \equiv St \, D/2$,
for various particle diameter $d_p$ at
$Re = 1132$ with $D = 1.5$.
Because $U > 10$ m/s 
and therefore the value of Froude number $Fr \equiv U^2/(g D)$
is greater than $6795$, the effect of gravity is 
expected to be negligible according to \cite{huang2001} 
although gravity is included in the OpenFOAM code by default
in (\ref{f_d}).

Although in typical impactors for particle size analysis 
the jet-to-plate distance $S$ is comparable to the nozzle size $D$,
very large $S/D$ (e.g., $S/D \sim 10$) is usually employed in the 
nozzle-to-plate settings with  
Aerosol Jet$^{\circledR}$ 
printing.
Therefore, the present study examines a range of $S/D$ from
$0.5$ to $4$ for gaining insights into both traditional inertial impactor 
and 
Aerosol Jet$^{\circledR}$ 
applications.

\subsection{Straight nozzle without taper ($\phi = 0$)}
The case without nozzle taper represents 
the simplest geometric configuration for computational modeling,
and therefore is especially of theoretical importance 
because aerosol flow in the nozzle channel is well defined and 
straightforward.

\begin{figure}[t!] \label{pattern1}
\includegraphics[scale=0.5]{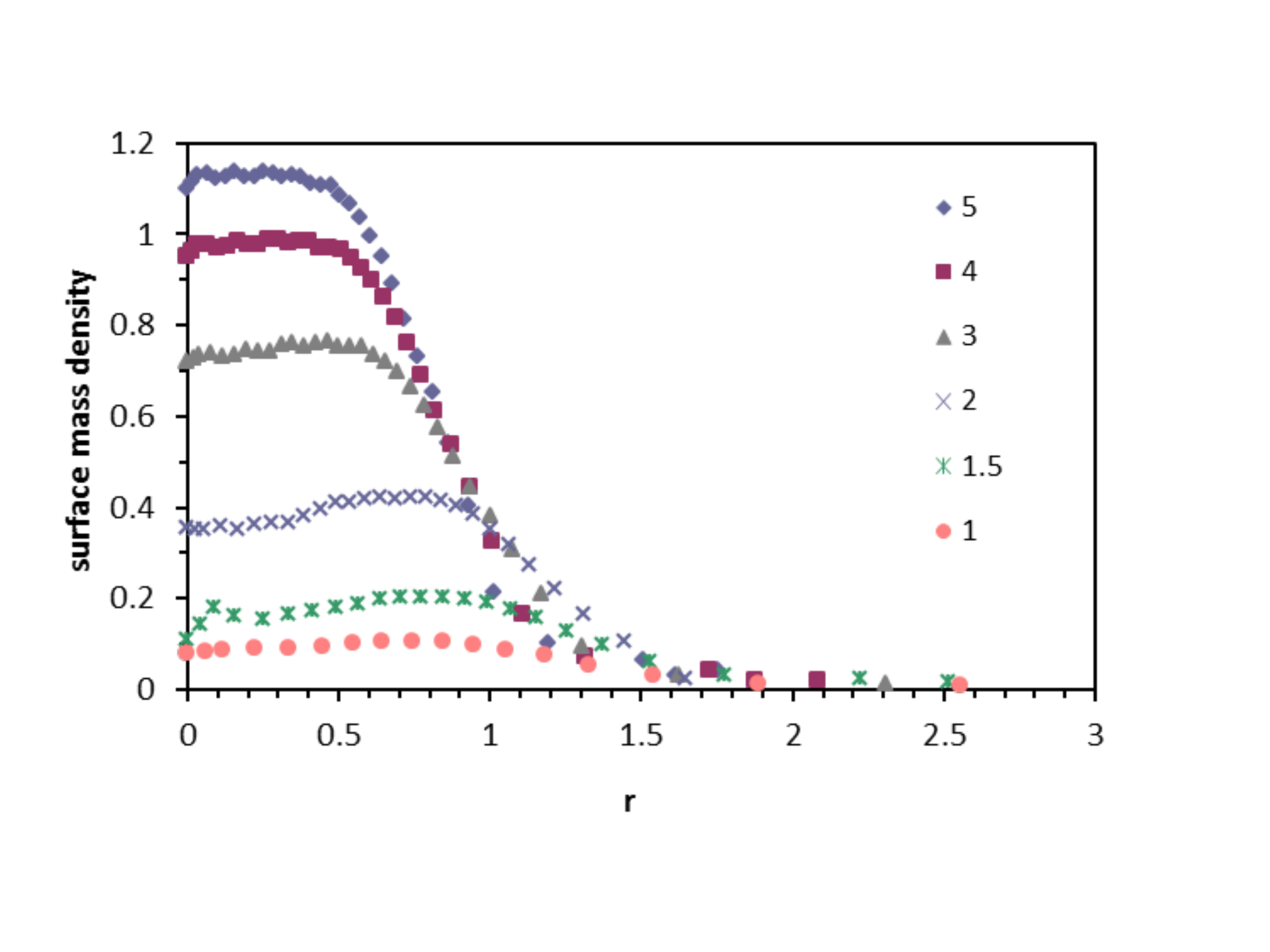}
\caption{
The dimensionless particle density 
$\sigma$ as a function of normalized radial position $r$
(in units of nozzle radius $D/2 = 0.75$ mm) on the
impaction plate 
for $Re = 1132$ with $D = 1.5$ mm, $S/D = 1$, $T/D = 2$, and $\phi = 0^o$. 
The labels are the values of particle diameters $d_p$ in units of $\mu$m   
with particle density of $\rho_p = 1000$ kg m$^{-3}$. The corresponding
values of $\sqrt{St}$ are $1.096$, $0.881$, $0.665$, 
$0.449$, $0341$, and $0.233$, as given in Table 1.}
\end{figure}

Fig 3 shows particle deposition patterns in terms of 
the dimensionless particle density 
$\sigma$ as a function of normalized radial position $r$ on the
impaction plate, determined according to 
(\ref{p-density}).
In this case,
the jet Reynolds number $Re$ is $1132$ (for a flow rate of $Q = 1200$ ccm
through a nozzle of $D = 1.5$ mm).
With particle density of $\rho_p = 1000$ kg m$^{-3}$, 
the particles of diameters in a range of $0.5$ $\mu$m to $5$ $\mu$m
cover the range of $\sqrt{St}$ from $0.1244$ to $1.0966$ (cf. Table 1),
wherein most significant variation of 
the particle deposition efficiency $\eta$ is expected.
Obviously, fewer particles of relatively smaller size 
could be deposited on the impaction plate; 
therefore, we have relatively lower surface mass density $\sigma$ 
for smaller $\sqrt{St}$.
It is interesting to note the particle deposition pattern, 
or profile of $\sigma(r)$,
also changes with the value of particle size $d_p$ or $\sqrt{St}$, 
namely, smaller particles tends to have slightly lower deposition rate
at the center (around $r = 0$) with 
relatively higher deposition rate extending to
larger radial position $r$.
For larger particles such as $d_p = 4$ and $5$ $\mu$m 
or $\sqrt{St} = 0.8807$ and $1.0966$, 
the total deposited mass are the same with $\eta = 100\%$ (cf. Table 2)
but the particle deposition patterns exhibit noticeable differences.
The profile of $d_p = 4$ $\mu$m is about $10\%$ lower than 
that of $5$ $\mu$m in the middle region ($r < 1$), 
but becomes higher at larger radial distance ($r > 1$) 
such that the two different profiles corresponds to the same
amount of normalized particle volume on the impaction plate.

\begin{figure}[t!] \label{pattern2}
\includegraphics[scale=0.34]{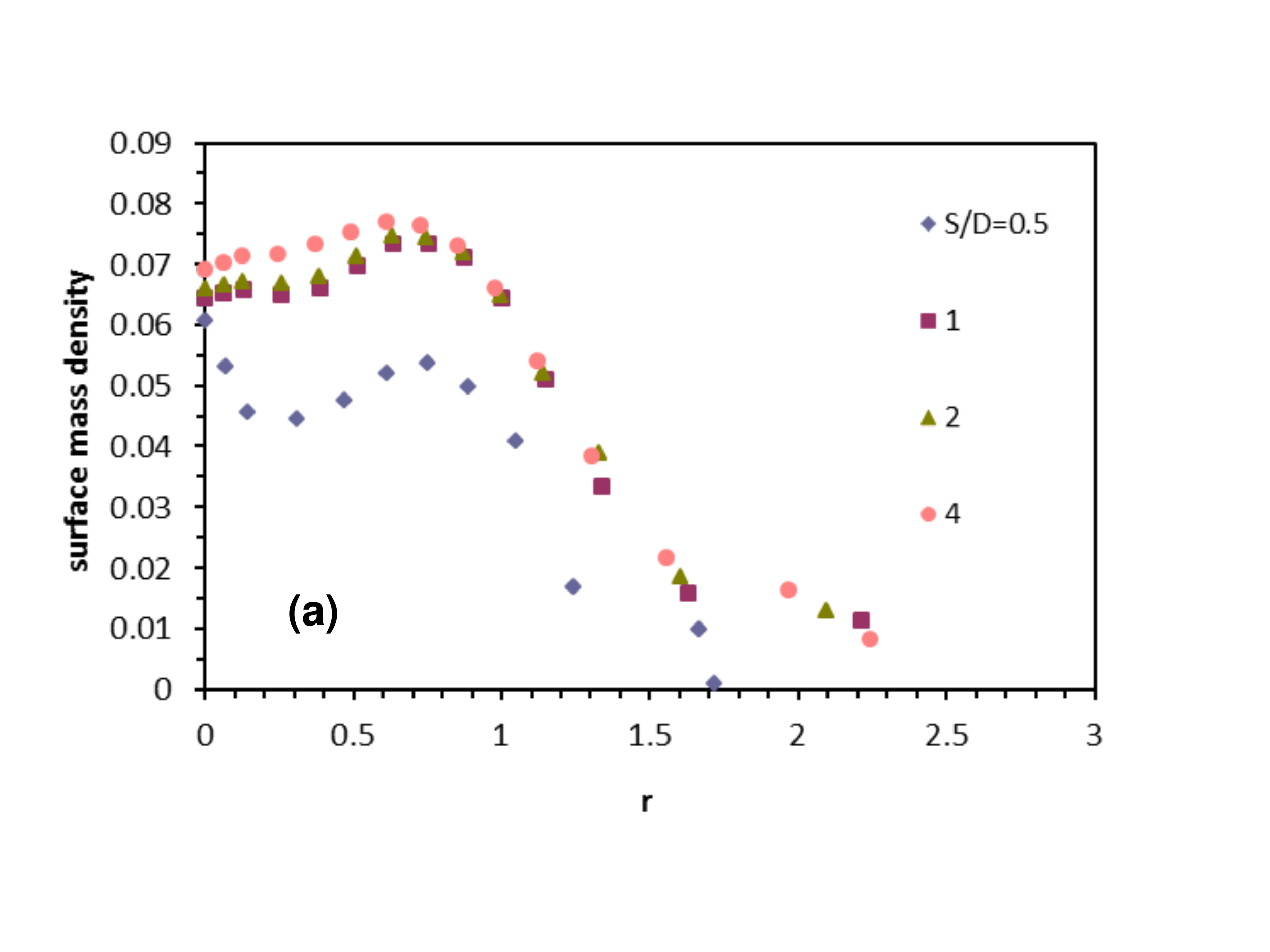}
\includegraphics[scale=0.34]{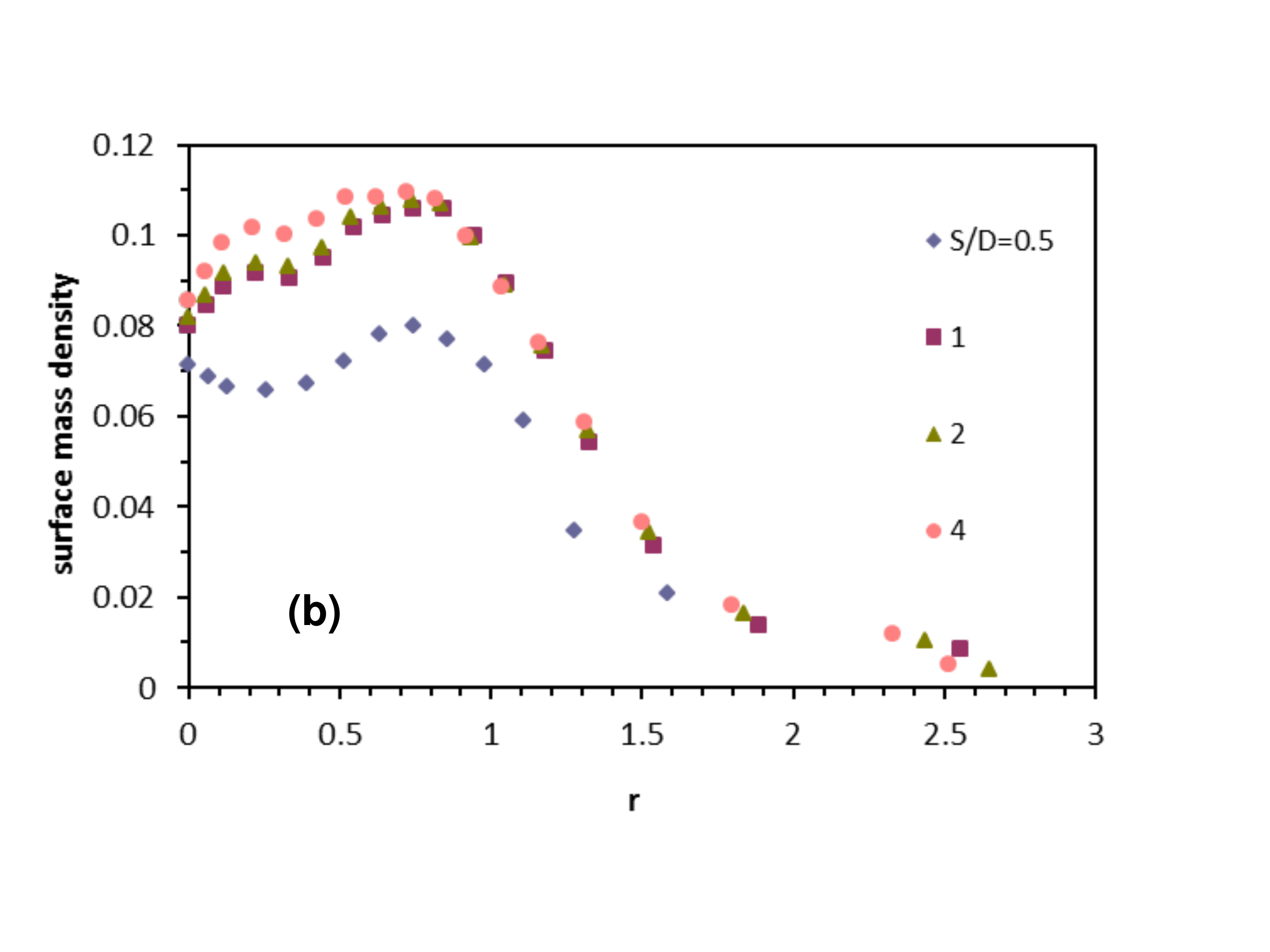}
\includegraphics[scale=0.34]{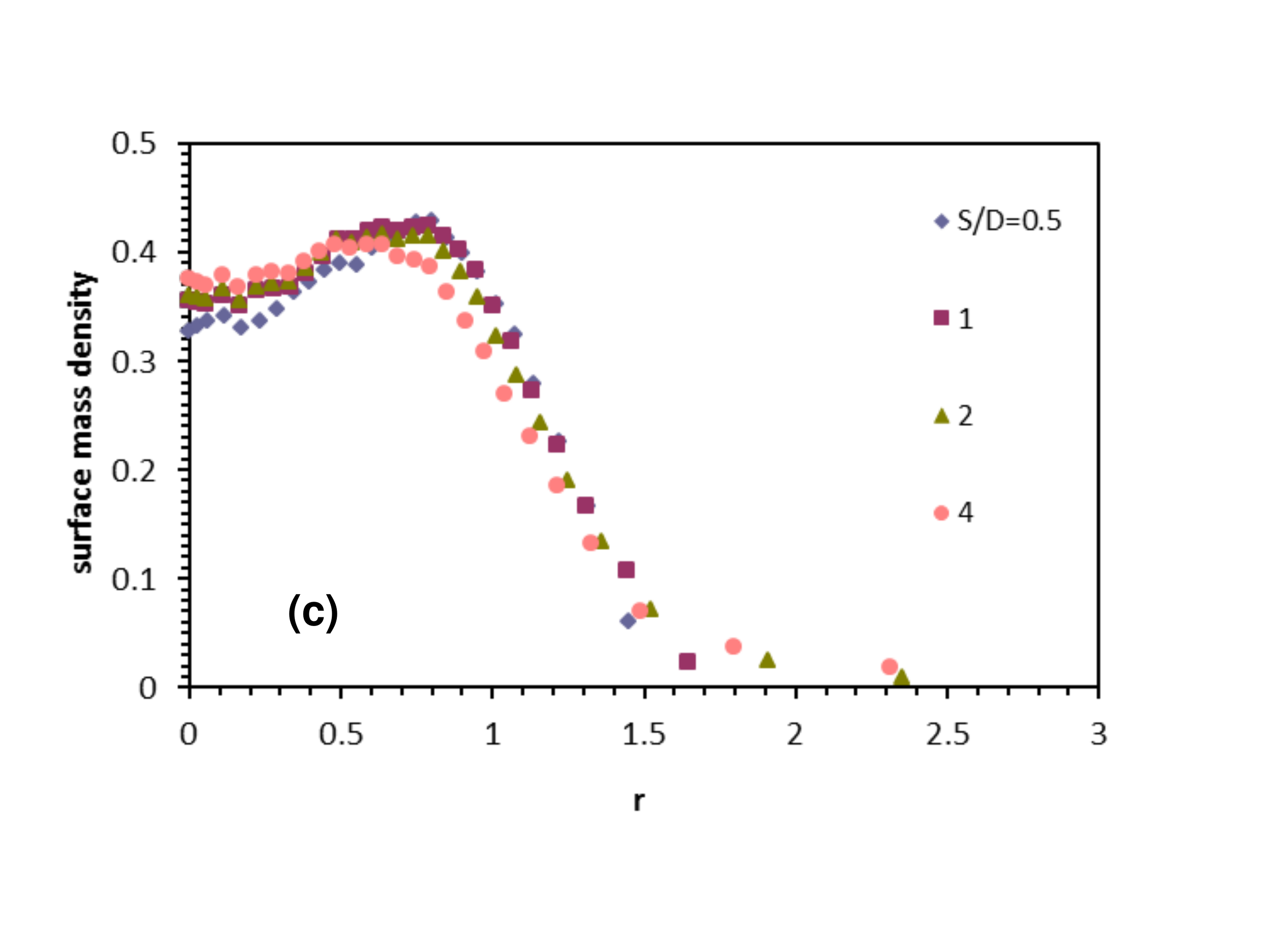}
\includegraphics[scale=0.34]{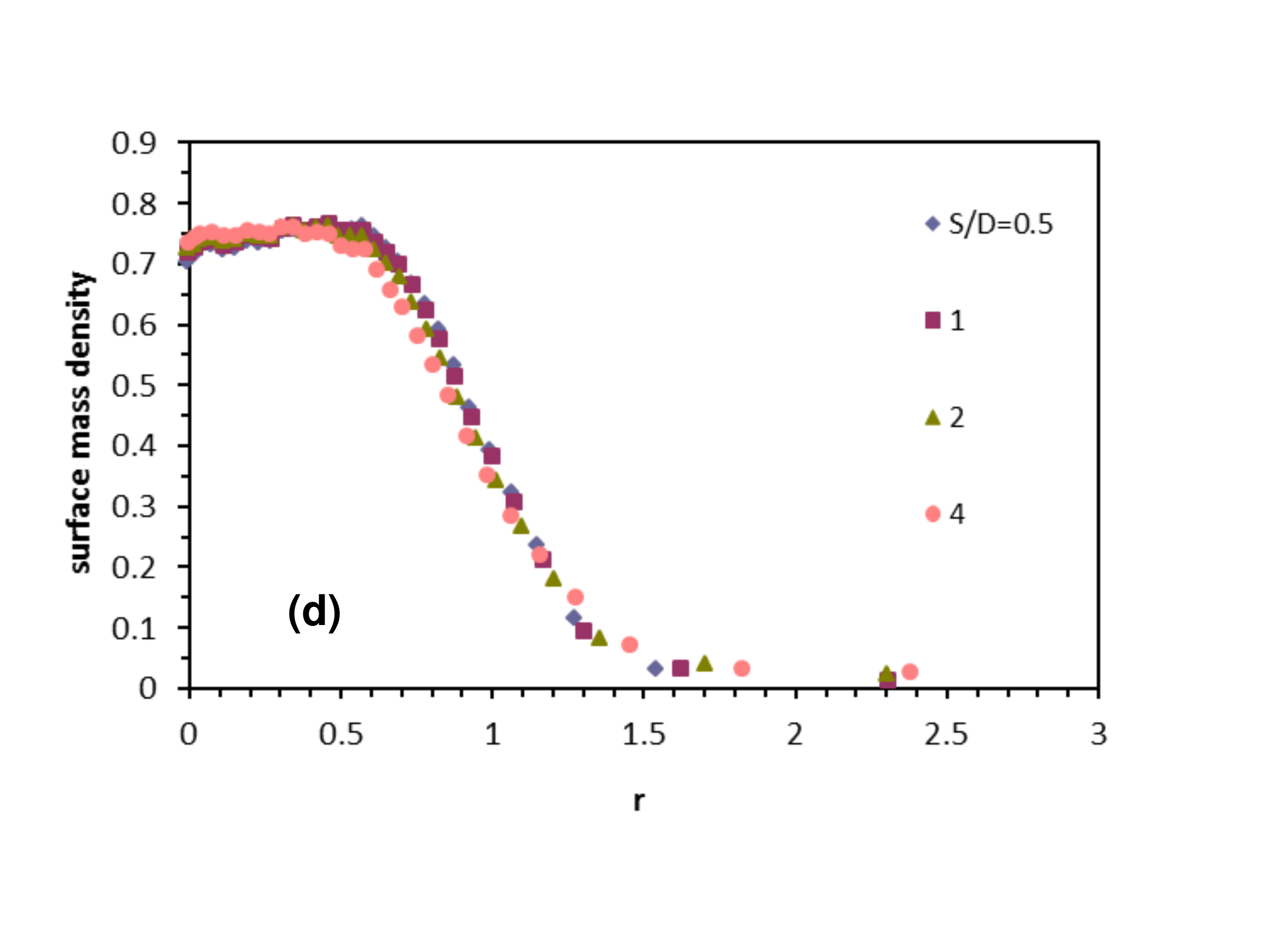}
\caption{As in Fig. 3 but 
with $S/D = 0.5$, $1$, $2$, and $4$ as labeled.
Plot (a) is for $d_p = 0.5$ $\mu$m ($\sqrt{St} = 0.124$), 
(b) $d_p = 1$ $\mu$m ($\sqrt{St} = 0.233$),   
(c) $d_p = 2$ $\mu$m ($\sqrt{St}=0.449$), 
and (d) $d_p = 3$ $\mu$m ($\sqrt{St} = 0.665$).}
\end{figure}

Because the particle deposition patterns typically 
exhibit a rapid declining edge (followed by a rather 
long tail with insignificant amount of particle deposition),
it may be meaningful to use
the radial position for`surface mass density $\sigma$
to reach half of its peak value (where the slope is 
expected to be close to its maximum magnitude as can be easily determined
in profilometer measurements)
to define the deposition spot size, namely,
the ``spot radius'' $\bar{r}$ 
(or ``spot diameter'' $\bar{d}$ $\equiv 2 \, \bar{r}$). 
For example, the cases of $d_p = 5$ $\mu$m ($\sqrt{St} = 1.0966$) 
and $d_p = 3$ $\mu$m ($\sqrt{St} = 0.6648$)
in Fig. 3 have $\bar{r} \approx 0.79$ and $\approx 0.93$, respectively 
(both are $< 1$, i.e., less than nozzle radius). 
But for $d_p = 2$ $\mu$m ($\sqrt{St} = 0.4489$) 
and $d_p = 1$ $\mu$m ($\sqrt{St} = 0.2328$), 
the spot radius 
would become $\bar{r} \approx 1.19$ and $1.24$, respectively (larger than
nozzle radius).
It is understandable that the particle deposition pattern 
spreads out as indicated by the increasing value of $\bar{r}$ 
with reducing $\sqrt{St}$,
because the smaller particles tend to follow the deflecting streamlines 
of gas flow more closely.
Interestingly, for the case of $d_p = 0.5$ $\mu$m 
($\sqrt{St} = 0.1244$, not shown in Fig. 3),
the value of $\bar{r}$ is $\approx 1.21$.
The reason for a shrinking $\bar{r}$ with further reduced $\sqrt{St}$ is 
that particles located in larger rings away from the axis 
(having larger $\hat{r}_i$) are unable to impact the 
plate but to be carried by the deflecting gas flow to the outlet.

Fig. 4 shows the effects of varying the (normalized) ``jet-to-plate''
distance $S/D$ with the straight nozzle at $Re = 1132$ for 
$\sqrt{St} = 0.124$ (a), $0.233$ (b), $0.449$ (c), and
$0.665$ (d). 
The deposition pattern varies considerably with changing of $S/D$ 
from $0.5$ to $4$
for small particles 
with $\sqrt{St} < 0.4$, but becomes relatively insensitive to
variation of $S/D$ when $\sqrt{St}$ is greater than $0.5$.
The change in particle deposition pattern is mostly noticeable 
when $S/D$ varies between $0.5$ and $1$, especially for small $\sqrt{St}$,
but not much so for $S/D > 1$.
The values of spot radius $\bar{r}$ in Fig. 4(a) $\sqrt{St} = 0.124$ 
are $\approx 1.16$, $1.21$, $1.21$, and $1.20$ 
repectively for $S/D = 0.5$,
$1$, $2$, and $4$, 
whereas that in Fig. 4(b) $\sqrt{St} = 0.233$ 
become $\bar{r} \approx 1.16$, $1.24$, $1.24$, and $1.24$.
Thus, the spot radius seems to increase slightly with increasing jet-to-plate
distance $S/D$, when $\sqrt{St}$ is small (e.g., $< 0.4$).
However, for the case of Fig. 4 (c) $\sqrt{St} = 0.449$,
we have $\bar{r} \approx 1.20$, $1.19$, $1.19$, and $1.15$
repectively with $S/D = 0.5$,
$1$, $2$, and $4$.
Similarly, the case of Fig. 4 (d) $\sqrt{St} = 0.665$ has
$\bar{r} \approx 0.95$, $0.93$, $0.91$, and $0.89$ 
for $S/D = 0.5$,
$1$, $2$, and $4$, respectively.
The spot radius tends to decrease slightly with increasing $S/D$ 
for relatively large $\sqrt{St}$.

Table 2 shows computed deposition efficiancy $\eta$ 
for various particle sizes $\sqrt{St}$ 
with $S/D = 0.5$, $1$, $2$, and $4$ labled as the subscript of $\eta$.
Again, the change in deposition efficiency is mostly noticeable 
when $S/D$ varies from $0.5$ to $1$ especially for $\sqrt{St} < 0.4$, 
but not much so for $S/D > 1$.
It is a bit counterintuitive to find that the deposition efficiency 
for a given $\sqrt{St}$ tends to be lower for $S/D = 0.5$, 
namely, when the impaction plate is closer to the nozzle exit,
especially for relatively small $\sqrt{St}$ (with small particles).
This suggests an impactor with relatively smaller jet-to-plate distance
to yield a sharper cut for deposition efficiency in terms of particle size.
It seems that larger $S/D$ (e.g. $S/D = 4$) 
would lead to a more gradual change of $\eta$ versus $\sqrt{St}$ 
especially at relatively small $\sqrt{St}$.
The value of $\sqrt{St_{50}}$ 
seems to be rather insensitive to the change of $S/D$ 
in the range from $0.5$ to $4$, varying within $10\%$ from 
the average value $0.362$ with 
a minimum of $0.346$ around $S/D = 1$. 

\begin{table}
\caption{Impaction efficiency $\eta$ versus $\sqrt{St}$ 
for various particle diameter $d_p$ with $S/D = 0.5$, $1$, $2$, and $4$ 
(which is used as the subscript of $\eta$) at
$Re = 1132$ with $D = 1.5$ mm and taper half angle $\phi = 0$
(straight nozzle).
The values of $\sqrt{St_{50}}$ 
corresponding to $\eta = 0.5$ are $0.389$, $0.346$, $0.351$, and $0.362$
for $S/D = 0.5$, $1$, $2$, and $4$, respectrively.}
\begin{center}
\begin{tabular*}{0.75\textwidth}{@{\extracolsep{\fill}} lccccc}
\hline
\hline
\\ %\vspace{ 1 mm }
  $d_p$ ($\mu$m) & $\sqrt{St}$ & $\eta_{0.5}$ & $\eta_1$ & $\eta_2$ & $\eta_4$ \\
\hline
\\
  0.5 & $0.1244$ & $0.093$ & $0.172$ & $0.172$ & $0.185$ \\
  1 & $0.2328$ & $0.141$ & $0.259$ & $0.266$ & $0.266$ \\
  1.5 & $0.3409$ & $0.316$ & $0.486$ & $0.475$ & $0.455$ \\
  1.75 & $0.3949$ & $0.523$ & $0.634$ & $0.613$ & $0.572$ \\
  2 & $0.4489$ & $0.711$ & $0.757$ & $0.734$ & $0.689$ \\
  3 & $0.6648$ & $0.929$ & $0.942$ & $0.942$ & $0.942$ \\
  4 & $0.8807$ & $0.981$ & $1.000$ & $1.000$ & $1.000$ \\
  5 & $1.0966$ & $1.000$ & $1.000$ & $1.000$ & $1.000$ \\ 
\hline
\hline
\end{tabular*}
\end{center}
\end{table}

Another observation from Table 2 is that 
significant amount of deposition still occurs (e.g., $\eta \sim 10\%$)
with particles of very small $\sqrt{St}$
($\sim 0.1$),
indicating an ever-present 
``small particle contamination'' on an impaction plate intended to
catch only larger particles.
If the value of the Stokes number $St$ is evaluated according to 
the ratio of stop distance and
the actual particle radial position from the nozzle axis at the
nozzle exit (instead of $D/2$), 
the particles closer to the axis (with smaller radial distance) 
would have larger effective values of $St$ and therefore 
would be expected to impact the plate.
Based on this logic,
the particles around axis should have very large effective $St$ 
and always impact the plate regardless their sizes.
Another way of explaining this phenomenon is to consider the radial 
component of laminar flow velocity $u_r$ as a circular jet impinging on 
a solid surface.
As the jet is approaching the surface, the flow spreads in the radial
direction as the axial velocity $u_z$ decreases. 
According to the continuity equation (\ref{continuity}), 
$u_r$ is expected to increase with $r$ starting from zero at $r = 0$;
in other words, $u_r \to 0$ as $r \to 0$. 
Thus, in the vicinity of the axis ($r = 0$), $u_r$ diminishes 
such that the 
deflecting flow parallel to the impaction wall tends to disappear
and particles near the axis always arrive at the plate 
without being deflected. 
As a check, computations are performed with
a reduction of $\rho_p$ from $1000$ to $10$ 
which effectively reduces $\sqrt{St}$ from $0.1244$ to $0.0124$ 
for $d_p = 0.5$ $\mu$m
(by an order of magnitude) and the results show  
$\eta_{0.5} = 0.078$ and $\eta_4 = 0.156$. 
Thus, about 
$10\%$ of fine particles with $0.01 \le \sqrt{St} \le 0.1$ 
are expected to always impact 
the plate, for $Re \sim 1000$.
In fact, the 
characteristic ``S'' shape deposition efficiency curves 
often observed experimentally with enhanced small particle impaction
had also been commented by
\cite{jurcik1995}
which seem to be consistent with their computational results.

\subsection{Tapered nozzle with $\phi = 15^o$}

\begin{table}
\caption{As in Table 2  
but for taper half angle $\phi = 15^o$ (tapered nozzle).
The values of $\sqrt{St_{50}}$ 
corresponding to $\eta = 0.5$ are $0.425$, $0.405$, $0.411$, and $0.427$
for $S/D = 0.5$, $1$, $2$, and $4$, respectrively.}
\begin{center}
\begin{tabular*}{0.75\textwidth}{@{\extracolsep{\fill}} lccccc}
\hline
\hline
\\ %\vspace{ 1 mm }
  $d_p$ ($\mu$m) & $\sqrt{St}$ & $\eta_{0.5}$ & $\eta_1$ & $\eta_2$ & $\eta_4$ \\
\hline
\\
  0.5 & $0.1244$ & $0.087$ & $0.119$ & $0.119$ & $0.119$ \\
  1 & $0.2328$ & $0.124$ & $0.167$ & $0.164$ & $0.158$ \\
  1.5 & $0.3409$ & $0.236$ & $0.294$ & $0.290$ & $0.286$ \\
  1.75 & $0.3949$ & $0.406$ & $0.467$ & $0.452$ & $0.420$ \\
  2 & $0.4489$ & $0.838$ & $0.787$ & $0.775$ & $0.645$ \\
  3 & $0.6648$ & $1.000$ & $1.000$ & $1.000$ & $1.000$ \\
  4 & $0.8807$ & $1.000$ & $1.000$ & $1.000$ & $1.000$ \\
  5 & $1.0966$ & $1.000$ & $1.000$ & $1.000$ & $1.000$ \\ 
\hline
\hline
\end{tabular*}
\end{center}
\end{table}

Many impactors are designed to have a tapered inlet 
for practical reasons \citep[cf.][]{marple1976}. 
In cascade impactors, the taper half angle is usually larger than
$15^o$. However, in the
Aerosol Jet$^{\circledR}$ 
printers the tapering channels
tend to have smaller $\phi$ for the desire of 
minimizing deviation of ink droplets from the flow streamlines.
Here, the effect of having a taper section in 
the nozzle inlet channel on particle deposition patterns 
is examined with the taper half angle specified as $\phi = 15^o$
(with $D = 1.5$ mm and $D_{in} = 3$ mm), 
among numerous possibilities.
The results are expected to adequately illustrate the general trends
of particle impaction behavior with tapered nozzle configuration.

\begin{figure}[t!] \label{pattern3}
\includegraphics[scale=0.5]{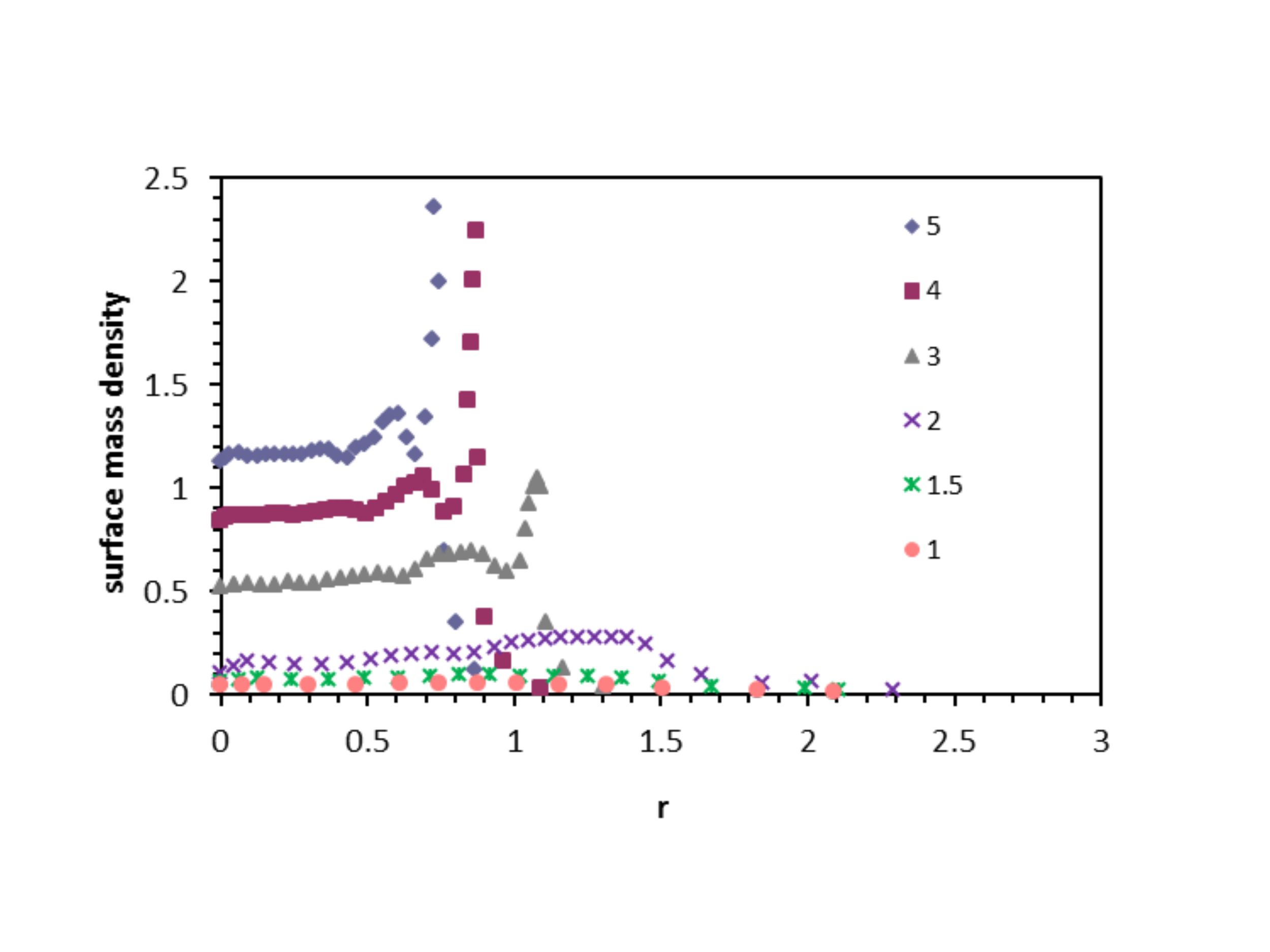}
\caption{As in Fig. 3 but 
for tapered nozzle with $\phi = 15^o$.} 
\end{figure}

Table 3 shows the values of computed
deposition efficiency $\eta$ versus $\sqrt{St}$ 
corresponding to various particle diameters with 
$S/D = 0.5$, $1$, $2$, and $4$, as in Table 2
but for cases with tapered nozzle having $\phi = 15^o$.
The value of $\eta$ seems to be rather insensitive to 
the change of $S/D$ for the most part, 
except that $\eta_4$ (for $S/D = 4$) becomes noticeably lower
than the others between $\sqrt{St} = 0.4$ and $0.6$.
The variation of $\sqrt{St_{50}}$ is also insigificant for 
different values of $S/D$ with an average value of $0.417$ and 
a minimun of $0.405$ around $S/D = 1$. 
Compared with the corresponding values of $\eta$ in Table 2 for
straight nozzle, the tappered nozzle tends to yield steeper slope 
of the deposition efficiency curve for a 
usually desired sharper ``cut''.  This could be 
a consequence of the aerodynamic focusing effect with the 
converging flow in tapered channel.
However, the particles flowing adjacent to the wall 
in a tapering channel 
can impact and stick on the channel wall due to their inertia.
For example, the present computations show that all particles 
of $d_p \ge 0.5$ $\mu$m ($\sqrt{St} \ge 0.1244$) 
placed within $\delta \sim 15$ $\mu$m from
the wall at inlet would impact on the nozzle channel wall 
instead of exiting the nozzle orifice.
This converts to about $2.0\%$ of the incoming particles with uniform 
concentration becoming ``wall loss'' inside the tapered nozzle channel. 
Such a wall loss is found to 
increase with increasing $\sqrt{St}$.
For particles of $\sqrt{St} = 1.0966$ ($d_p = 5$ $\mu$m, 
$\delta \sim 45$ $\mu$m), about $5.9\%$ 
would impact the tapered nozzle channel wall and become the wall loss. 
Therefore, the values of deposition efficiency $\eta$ in Table 3 are 
calculated based on fractions of particles actually exiting the nozzle
rather than fractions of 
incoming particles at inlet, 
by replacing $D_{in}$ in (\ref{eta}) with $D_{in} - 2 \delta$ 
where the value of $\delta$ depends on the particle size 
(generally increasing with $d_p$).
To prevent such a wall loss, introducing a coflowing 
sheath gas with a flow rate 
about $10\%$ of the particle laden flow may be practically feasible.

\begin{figure}[t!] \label{pattern4}
\includegraphics[scale=0.34]{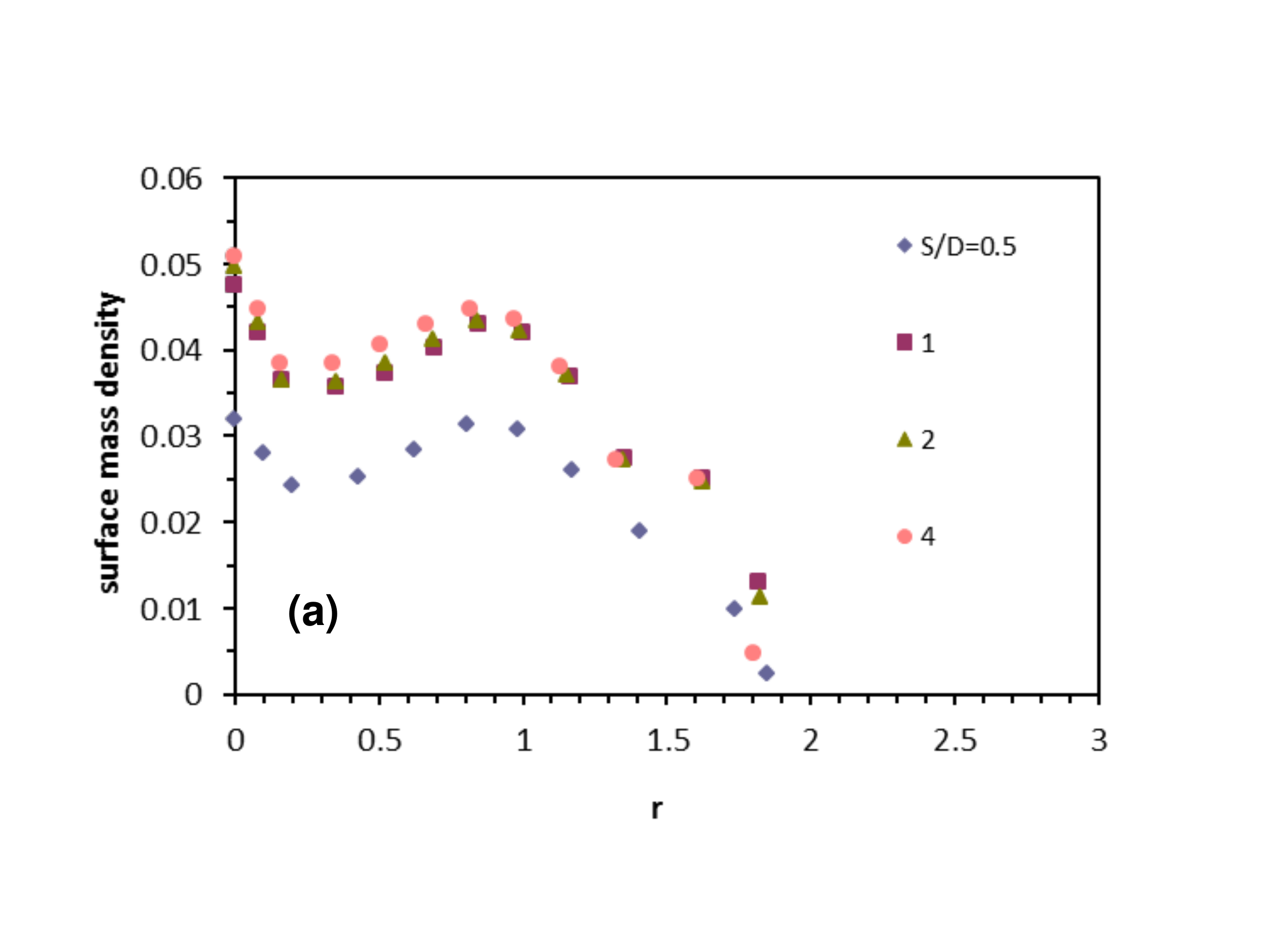}
\includegraphics[scale=0.34]{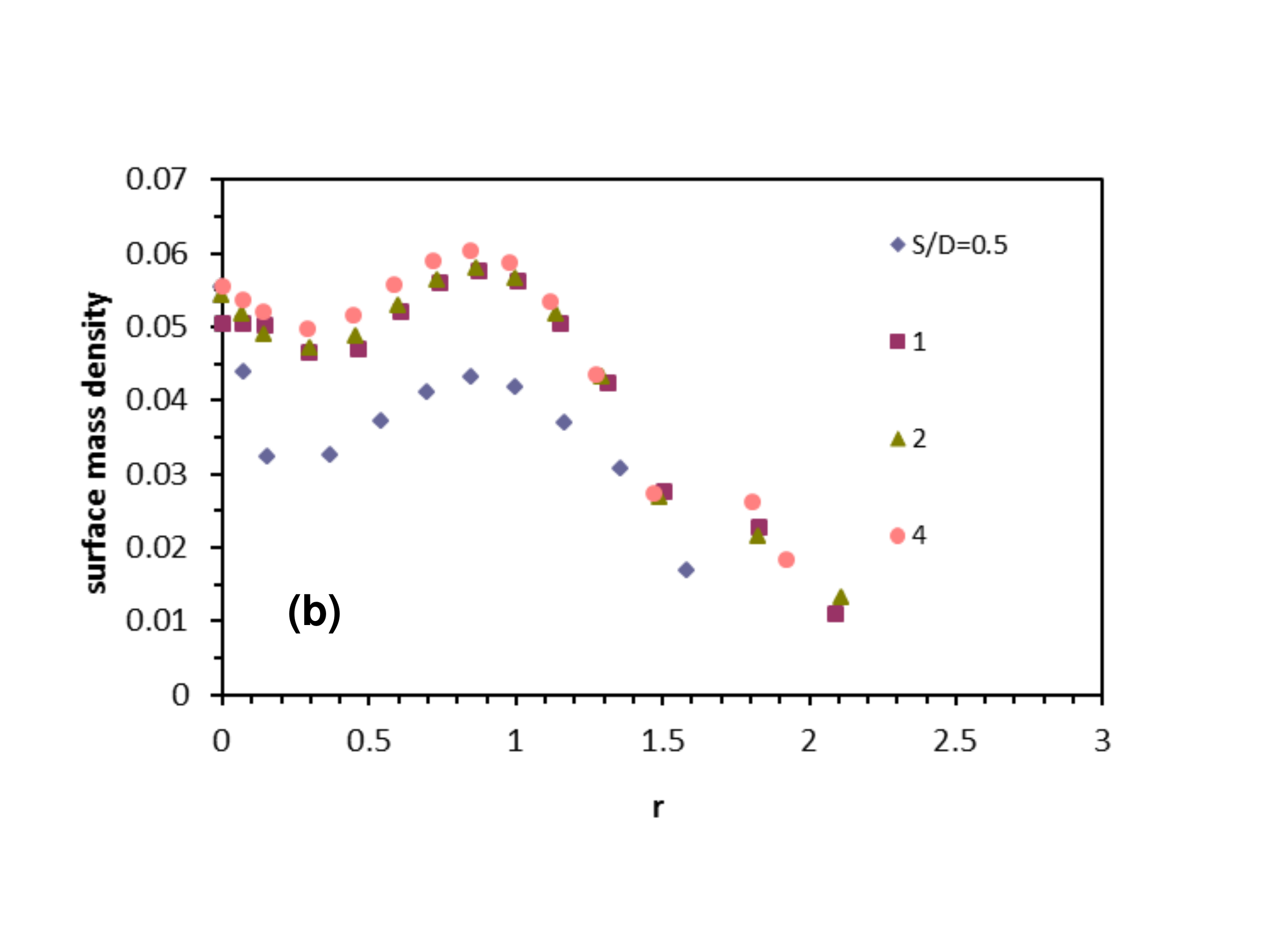}
\includegraphics[scale=0.34]{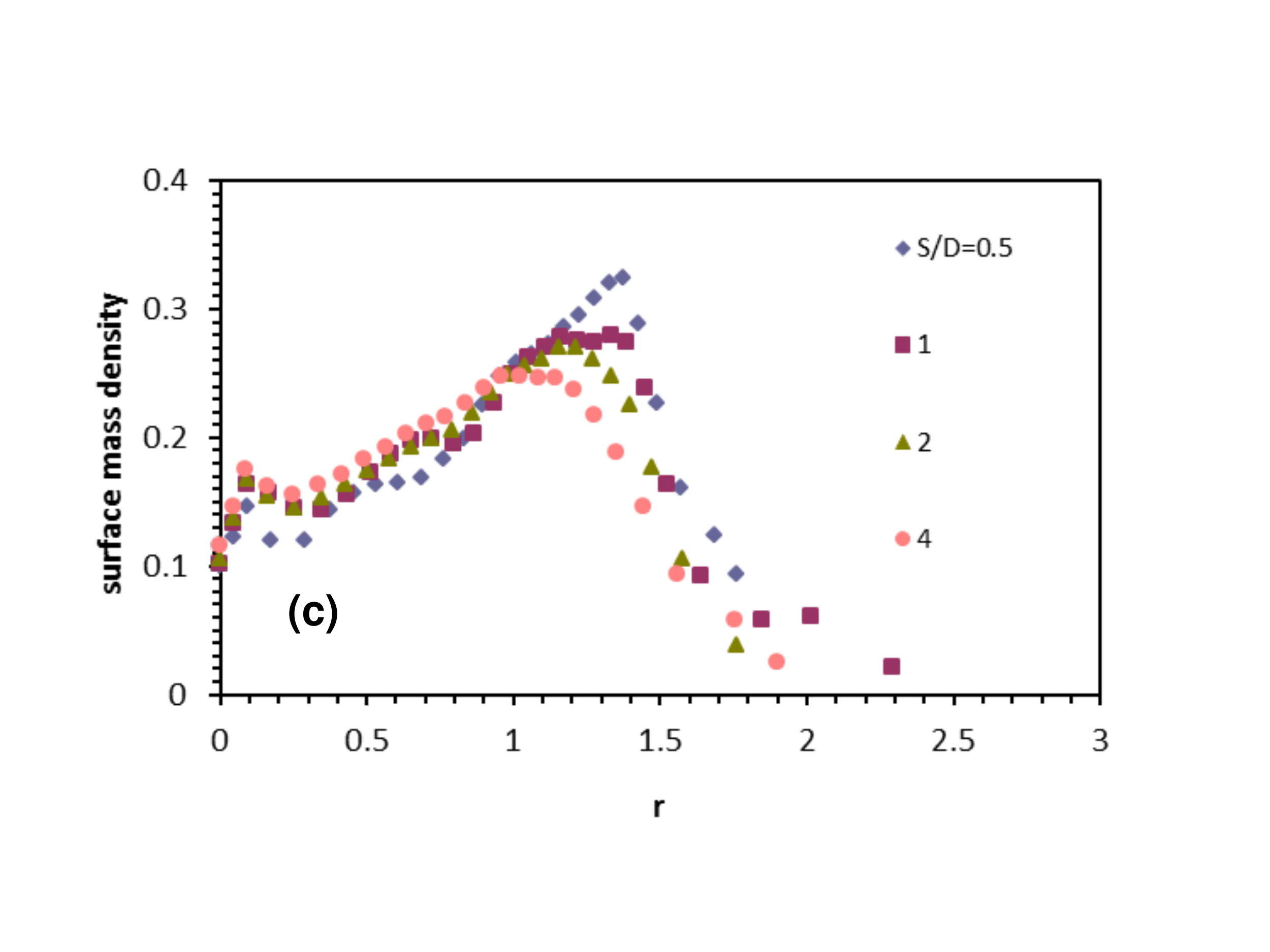}
\includegraphics[scale=0.34]{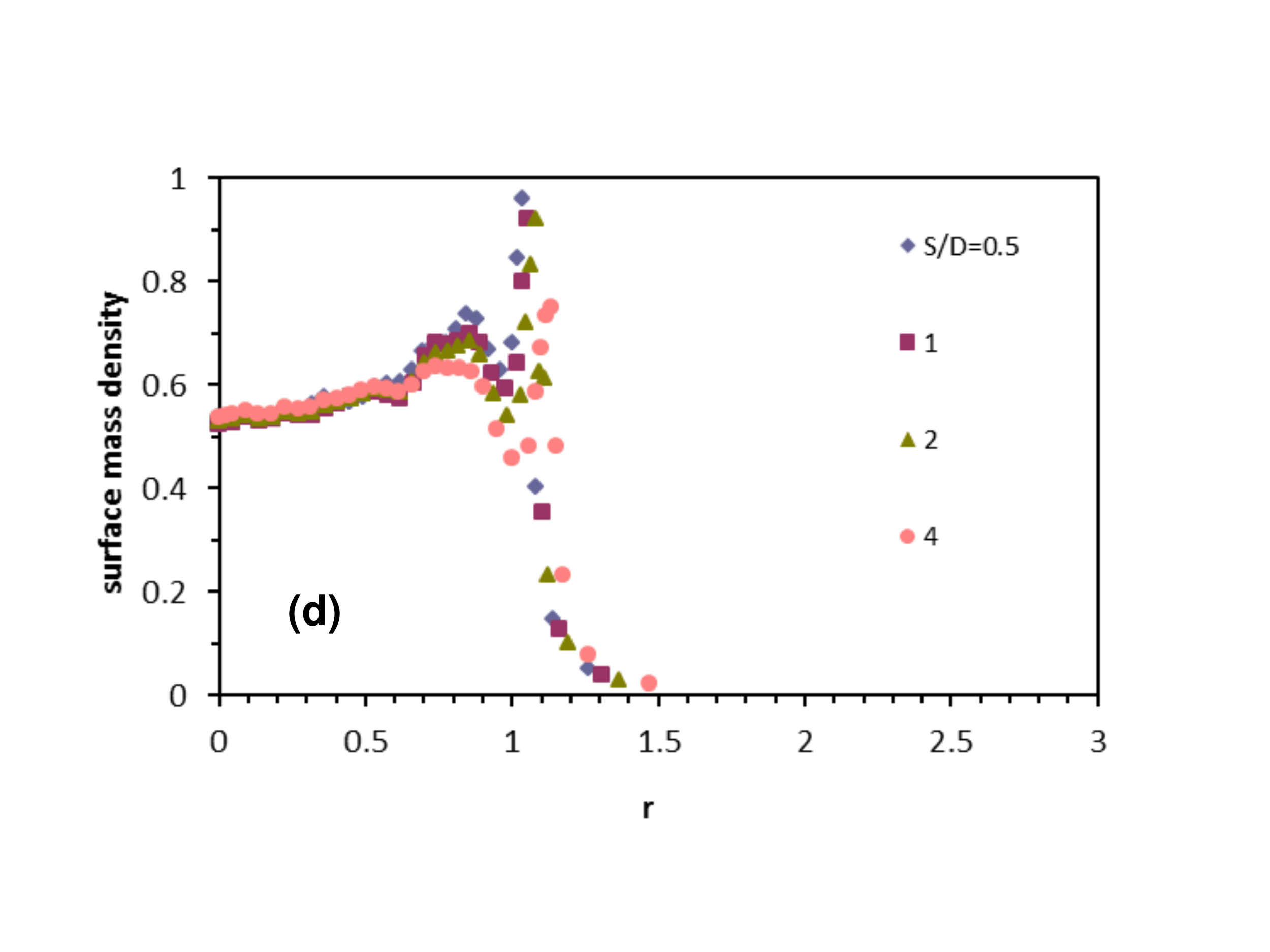}
\caption{As in Fig. 4 but 
for tapered nozzle with $\phi = 15^o$.} 
\end{figure}

As in Fig. 3 but for tapered nozzle with $\phi = 15^o$, 
profiles of dimenionless particle density as 
a function of $r$, namely $\sigma(r)$, on the impaction plate are 
shown in Fig. 5 for $d_p = 5$, $4$, $3$, $2$, $1.5$ and $1$ $\mu$m 
(as labeled). 
The differences made by tapering section appear to be quite significant,
when comparing Fig. 5 to Fig. 3. 
The converging flow in tapering channel induces the aerodynamic 
focusing effect that 
pushes particles inward \citep[cf.][]{dahneke1982, rao1993}, 
with stronger effect on 
larger particles further away from the axis 
(at larger $r$) than those closer to the axis.
Therefore, a region of concentrated particle impaction is expected 
in the particle deposition pattern.
Such a region of high particle density appears to be toward 
the edge of the circular deposition spot shown as a narrow sharp peaks 
in Fig. 5
for particles of relative large $\sqrt{St}$ ($> 0.6$ or $d_p \ge 3$ $\mu$m),
which is absent in Fig. 3.
Another noticeable effect of the tapered nozzle is 
that the ``spot radius'' $\bar{r}$ (defined in subsection 3.1) in Fig. 5
varies more substantially with $\sqrt{St}$ than that for straight nozzle
shown in Fig. 3, as expected based on the dependence of aerodynamic focusing 
effect upon particle size. 
For example, the values of $\bar{r}$ are $0.76$, $0.88$, $1.09$,
$1.52$, $1.52$, and $1.47$ for $\sqrt{St} = 1.0966$,
$0.8807$, $0.6648$, $0.4489$, $0.3409$, and $0.2328$, respectively 
(corresponding to the labels $5$, $4$, $3$, $2$, $1.5$, and $1$ in Fig. 5).
As their size decreases, the particles tends to follow
the outward deflecting gas flow more closely before impacting the plate, 
leading to generally increased values of $\bar{r}$. 
 
The effect of jet-to-plate distance $S/D$ on particle deposition
pattern is shown Fig. 6 for $S/D = 0.5$, $1$, $2$, and $4$ as labeled
with (a) $\sqrt{St} = 0.124$, (b) $0.233$, (c) $0.449$, and 
(d) $0.665$, as in Fig. 4 but for tapered nozzle with $\phi = 15^o$.
The profiles of $\sigma(r)$ for small particles, as shown in 
Figs. 6(a) and 6(b) for $\sqrt{St} = 0.124$ and $0.233$
have similar shapes for various $S/D$ except 
the ones for $S/D = 0.5$ are relatively lower as consistent with
relatively lower deposition efficiency $\eta$ in Table 3.
The values of $\bar{r}$ in Fig. 6(a) are $1.41$, $1.61$, $1.57$, and $1.55$
respectively for $S/D = 0.5$, $1$, $2$, and $4$, 
while those in Fig. 6(b) are $1.47$, $1.47$, $1.41$, and $1.36$.
The profiles in Fig. 6(c) for $\sqrt{St} = 0.449$ ($d_p = 2$ $\mu$m) 
show very similar ``donut'' shape (with a ring of high particle density 
toward deposition spot edge)
to that found experimentally by 
\cite{sethi1993} for their case of $\sqrt{St} = 0.48$, 
(which is the smallest $\sqrt{St}$ illustrated in their plots),
exhibiting a gradual piling peak toward the edge of 
the deposition spot. 
The values of $\bar{r}$ in Fig. 6(c) are $1.52$, $1.52$, $1.49$, and $1.45$
respectively for $S/D = 0.5$, $1$, $2$, and $4$,
while those for Fig. 6(d) for $\sqrt{St} = 0.665$
are $1.08$, $1.09$, $1.11$, and $1.16$.
The profiles of $\sigma(r)$ for larger $\sqrt{St}$ ($> 0.6$) 
also look similar to those illustrated by \cite{sethi1993} 
except their experimental data points are spaced a bit too coarse to
resolve the narrow peaks toward the spot edge.
Such a reasonable comparison with the experimentally measured 
pattern by \cite{sethi1993} 
may serve as an independent (qualitative) 
validation-varification for the present 
computational results.
The plots (c) and (d) in Fig. 6 exhibit a general trend of reducing 
peak particle density toward 
the deposition spot edge with increasing the jet-to-plate 
distance.

\subsection{Cases with $Re = 283$}

If all the parameters are kept the same except the flow rate at the inlet is 
reduced to $Q = 300$ ccm (from the nominal value of $1200$ ccm),
the value of the jet Reynolds number becomes $Re = 283$ (instead of
$1132$). Thus, the values of $\sqrt{St}$ for given $d_p$ 
become about one half of those in 
Tables 1--3, for convenience of comparison. 

\begin{table}
\caption{As in Table 2  
but for taper half angle $\phi = 15^o$ (tapered nozzle)
at $Re = 283$.
The values of $\sqrt{St_{50}}$ 
corresponding to $\eta = 0.5$ are $0.422$, $0.429$, $0.458$, and $0.500$
for $S/D = 0.5$, $1$, $2$, and $4$, respectrively.}
\begin{center}
\begin{tabular*}{0.75\textwidth}{@{\extracolsep{\fill}} lccccc}
\hline
\hline
\\ %\vspace{ 1 mm }
  $d_p$ ($\mu$m) & $\sqrt{St}$ & $\eta_{0.5}$ & $\eta_1$ & $\eta_2$ & $\eta_4$ \\
\hline
\\
  0.5 & $0.0622$ & $0.032$ & $0.053$ & $0.053$ & $0.047$ \\
  1 & $0.1164$ & $0.037$ & $0.062$ & $0.058$ & $0.050$ \\
  2 & $0.2245$ & $0.057$ & $0.090$ & $0.082$ & $0.074$ \\
  3 & $0.3324$ & $0.132$ & $0.190$ & $0.172$ & $0.137$ \\
  4 & $0.4404$ & $0.573$ & $0.537$ & $0.441$ & $0.323$ \\
  5 & $0.5483$ & $0.933$ & $0.893$ & $0.793$ & $0.642$ \\ 
  6 & $0.6562$ & $0.973$ & $0.959$ & $0.919$ & $0.919$ \\ 
  8 & $0.8702$ & $1.000$ & $1.000$ & $1.000$ & $1.000$ \\ 
\hline
\hline
\end{tabular*}
\end{center}
\end{table}

Shown in table 4 are values of the deposition efficiency $\eta$ 
versus $\sqrt{St}$ for various particle diameter $d_p$ 
with $S/D = 0.5$, $1$, $2$, and $4$ (as subscript of $\eta$) 
at $Re = 283$ with $\phi = 15^o$.
Comparing with those in table 3 at $Re = 1132$,
the values of $\eta$ corresponding to similar $\sqrt{St}$ are 
generally lower for $Re = 283$, as expected 
from less impinging momentum with thicker viscous boundary layer 
due to smaller $Re$.
Unlike the case with $Re = 1132$ in table 3, 
the value of $\sqrt{St_{50}}$ at $Re = 283$ in table 4 
appears to increase monotonically with $S/D$ 
from $\sqrt{St_{50}} = 0.422$ to $0.5$.
The values of $\eta$ indicate a trend of reduced sharpness of a cut 
for the deposition efficiency curve versus particle size when 
the jet-to-plate distance $S/D$ increases.

\begin{figure}[t!] \label{pattern5}
\includegraphics[scale=0.5]{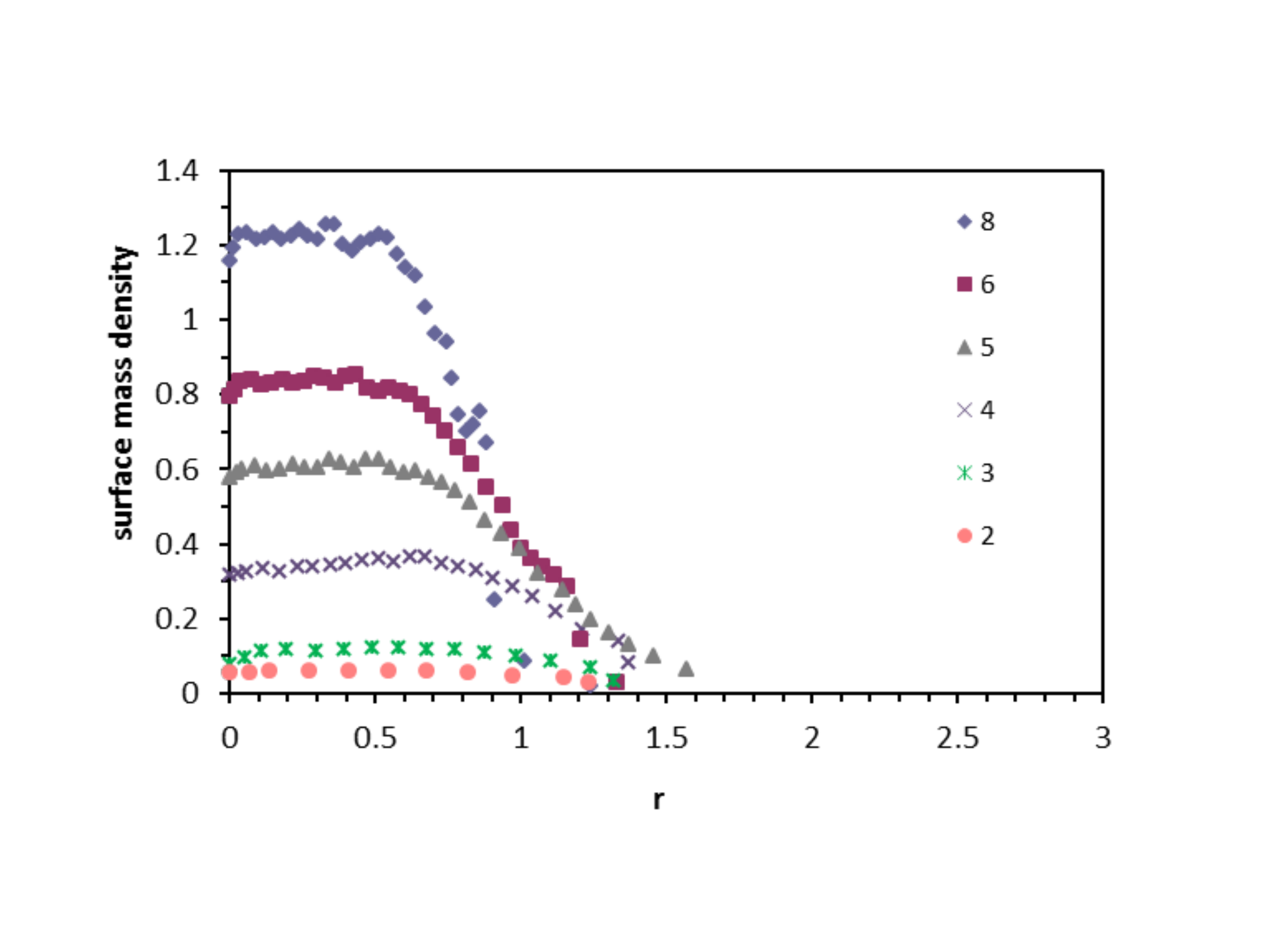}
\caption{As in Fig. 3 but 
for tapered nozzle with $\phi = 15^o$ at $Re = 283$ 
for $d_p = 8$ $\mu$m ($\sqrt{St} = 0.872$), $6$ $\mu$m ($0.656$),
$5$ $\mu$m ($0.548$), $4$ $\mu$m ($0.440$), 
$3$ $\mu$m ($0.332$), and $2$ $\mu$m ($0.224$).} 
\end{figure}

Similar to Figs. 3 and 5, 
the particle deposition pattern in terms of the
dimensionless particle density $\sigma$ 
is shown in Fig. 7 for tapered nozzle 
with $\phi = 15^o$ at $Re = 283$ 
as a function of radial position $r$ 
on the impaction plate.
In contrast to Fig. 5 with the same tapered nozzle (for $Re = 1132$),
the profiles of $\sigma(r)$ in Fig. 7 (for $Re = 283$)
do not show narrow high peaks 
toward edge even for relatively large $\sqrt{St}$;
they rather look similar to those profiles in Fig. 3 
for straight nozzle (at $Re = 1132$). 
It is interesting to note that 
the values of spot radius $\bar{r}$ (defined as the radius position on 
impaction plate for $\sigma(r)$ to reach half of its peak value, 
as defined above)
corresponding to similar values of $\sqrt{St}$ are 
fairly comparable between those in Fig. 7 and Fig. 3 (rather than Fig. 5).
For example,
corresponding to $\sqrt{St} = 0.872$, $0.656$, and $0.440$ in Fig. 7 
the values of $\bar{r}$ are estimated as
$0.76$, $0.88$, and $1.13$, respectively comparable to 
those in both Figs. 3 and 5. 
With smaller $\sqrt{St}$ at $0.224$ and $0.116$, 
the values of $\bar{r}$ become $\approx 1.19$ and $1.24$,
rather comparable to those in Fig. 3 but not Fig. 5.
Thus, with the same tapered nozzle 
at smaller jet Reynolds number $Re$ (e.g., $Re = 283$)
the high particle density peak toward
deposition spot edge disappears, while the particle deposition 
spot size does not change significantly especially for 
particles of $\sqrt{St} > 0.4$. 

\begin{figure}[t!] \label{pattern6}
\includegraphics[scale=0.34]{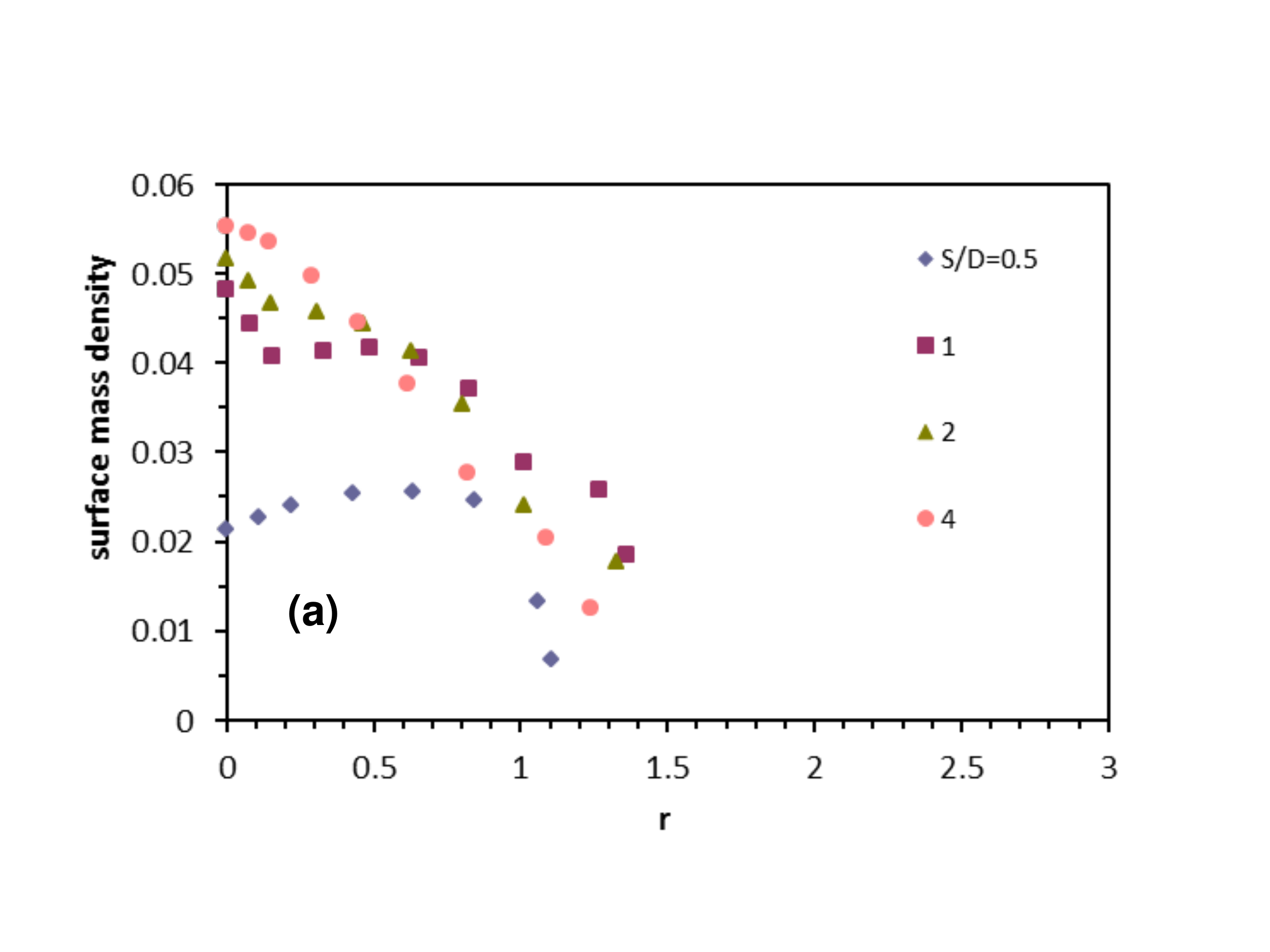}
\includegraphics[scale=0.34]{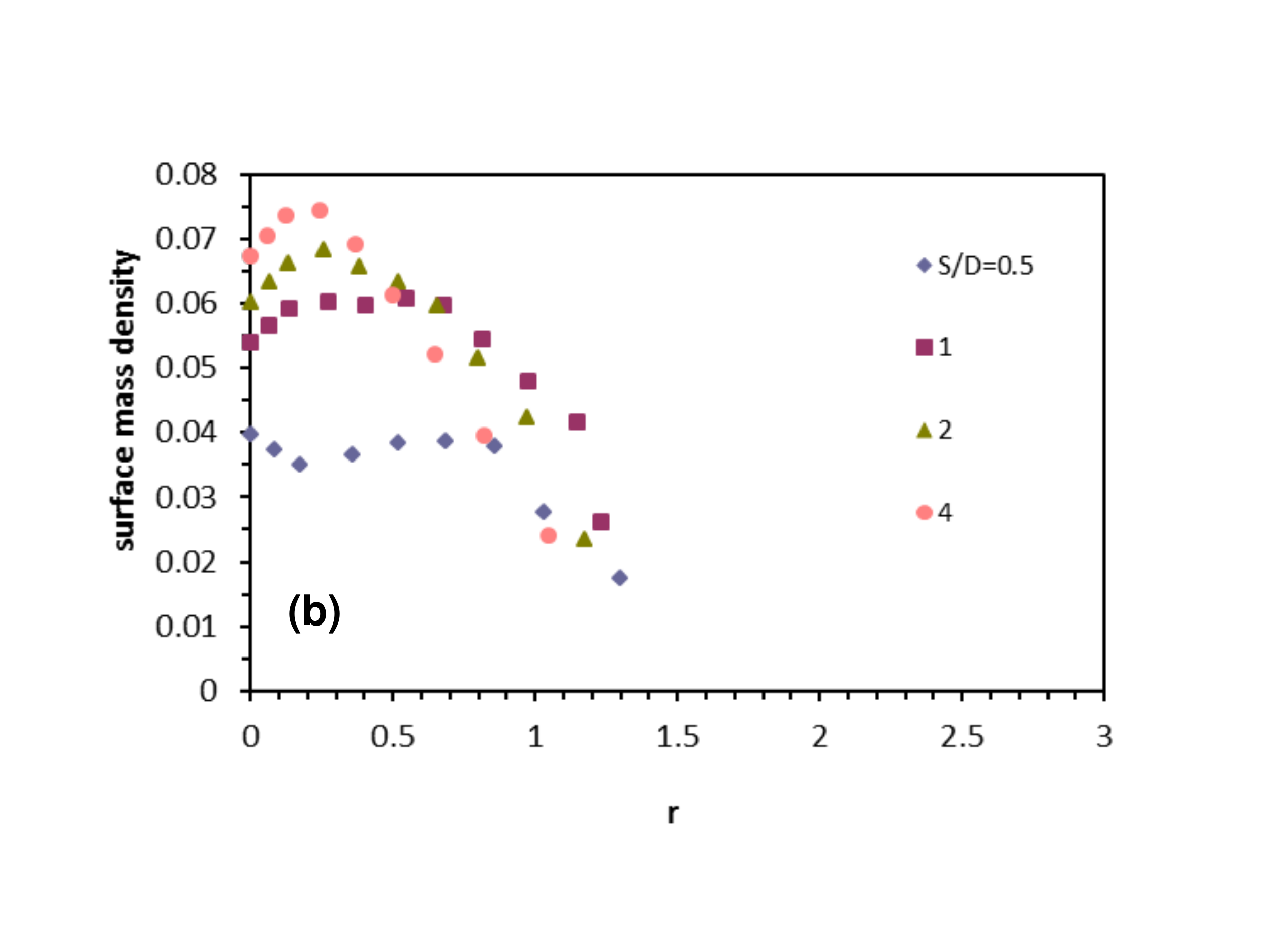}
\includegraphics[scale=0.34]{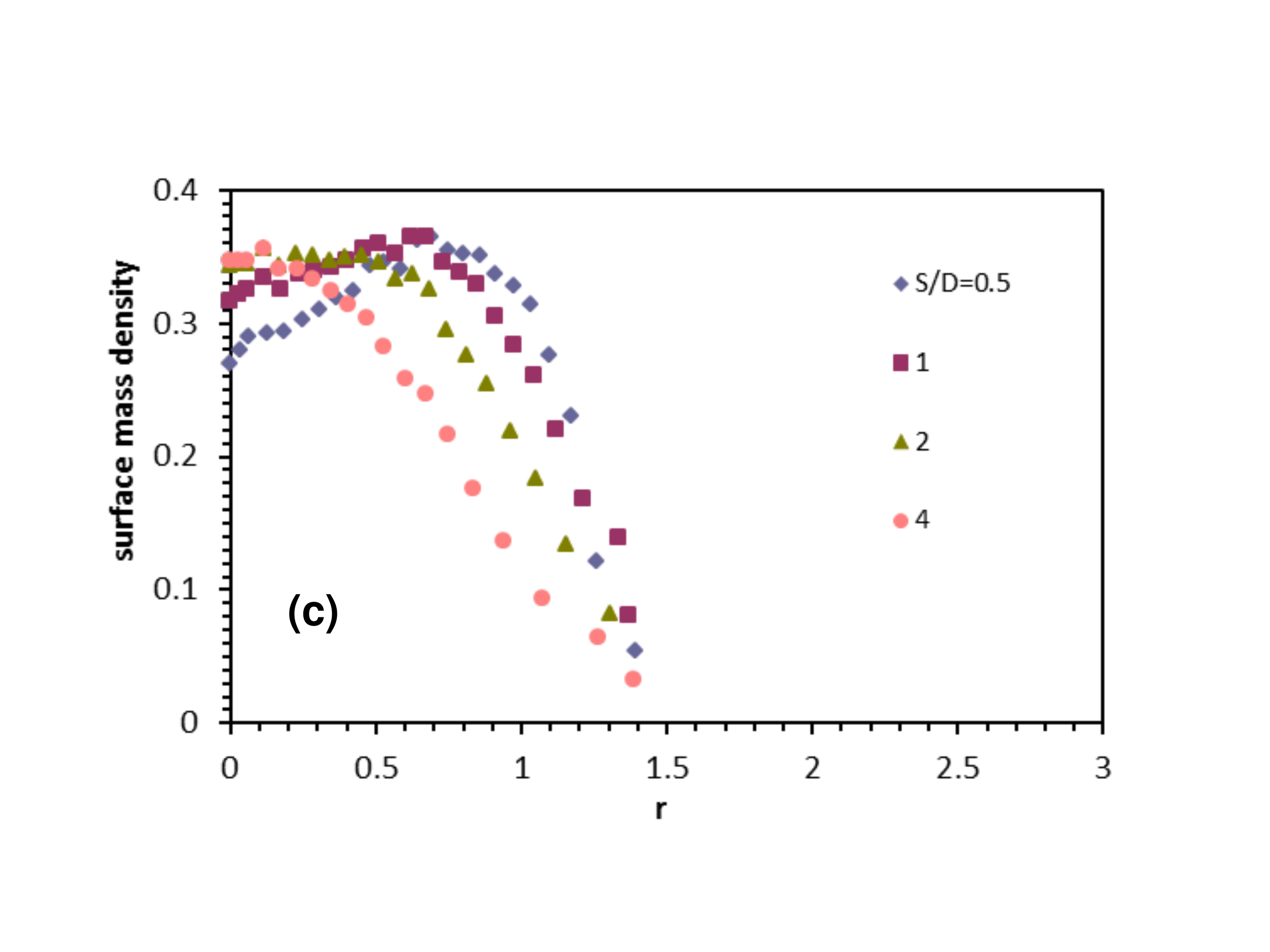}
\includegraphics[scale=0.34]{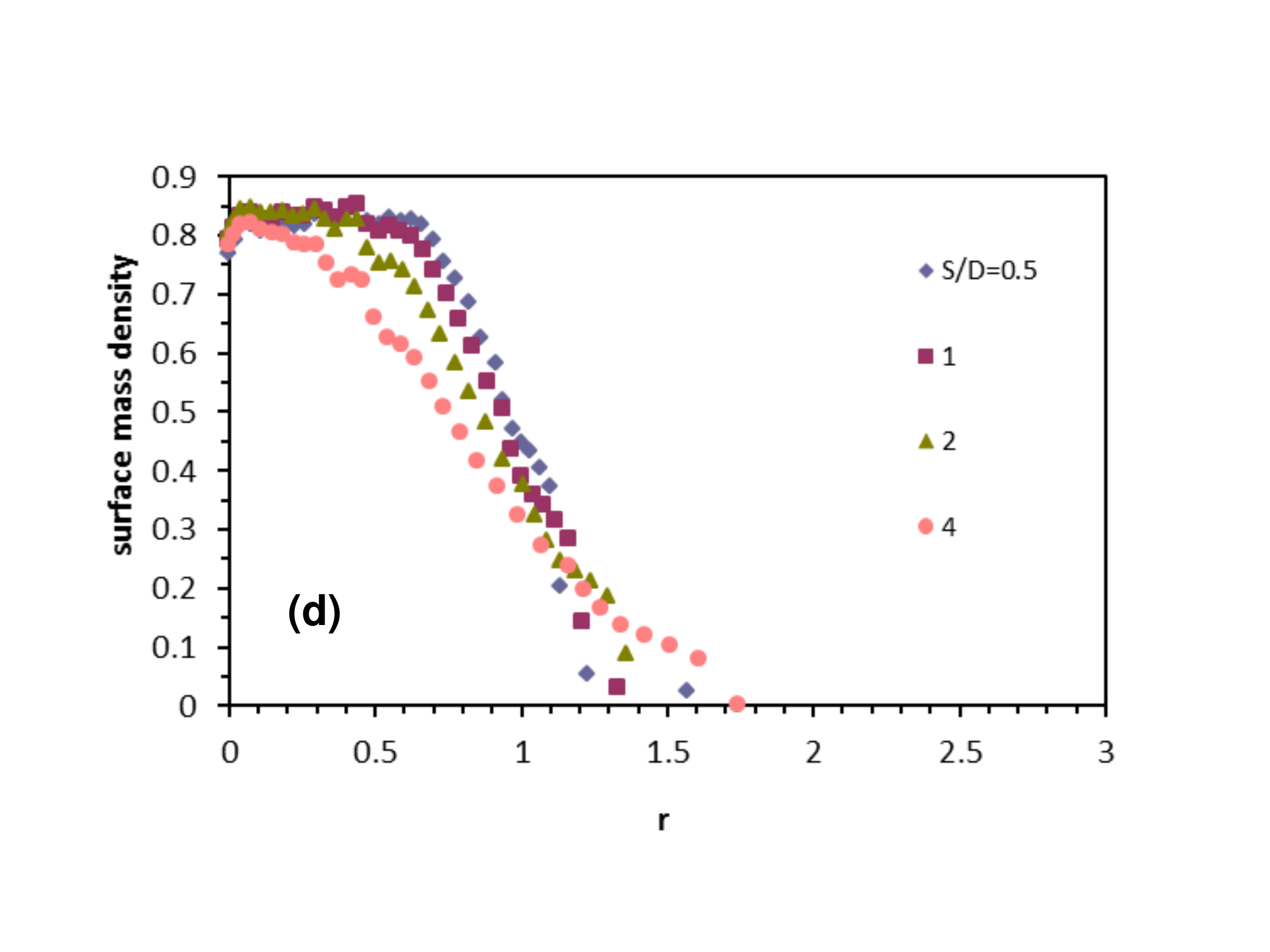}
\caption{As in Fig. 4 but 
for tapered nozzle with $\phi = 15^o$ at $Re = 283$. 
Plot (a) is for $d_p = 1$ $\mu$m ($\sqrt{St} = 0.116$),
(b) $d_p = 2$ $\mu$m ($\sqrt{St} = 0.224$), 
(c) $d_p = 4$ $\mu$m ($\sqrt{St} = 0.440$),
and (d) $d_p = 6$ $\mu$m ($\sqrt{St} = 0.656$).}
\end{figure}

The effect of jet-to-plate distance $S/D$ is shown in Fig. 8
for $Re = 283$.  
In contrast to those in Figs. 4 and 6 at similar values of $\sqrt{St}$,
the particle deposition pattern varies much more significantly  
with the change of the value of $S/D$ in Fig. 8,
which appears also consistent with the deposition efficiency data in Table 4.
For example, relatively much lower $\eta_{0.5}$ at small $\sqrt{St}$ 
is quite obvious 
for the case of $S/D = 0.5$ 
in Figs. 8(a) and 8(b). 
The decreasing $\eta$ with increasing $S/D$ at 
$\sqrt{St} = 0.440$ is clearly reflected in Fig. 8(c) 
with shrinking $\bar{r}$ ($\approx 1.19$, $1.13$, $1.01$, and $0.81$ 
for $S/D = 0.5$, $1$, $2$, and $4$, respectively).
However, at $\sqrt{St} = 0.656$ Fig. 8(d) shows 
distinctive difference between $\sigma(r)$ in the case of $S/D = 2$ and $4$,
with more gradual decline of $\sigma$ with $r$ (lack of edge definition)
for $S/D = 4$ ($\bar{r} \approx 0.79$) than that 
for $S/D = 2$ ($\bar{r} \approx 0.84$)
despite both have the same value of $\eta$ (cf. Table 4).
Such a feature of 
gradual declining $\sigma(r)$ without a clearly defined edge 
(as usually indicated by a sudden change of slope) 
becomes a common particle deposition pattern exhibited for cases 
with straight nozzle ($\phi = 0$) at $Re = 283$.
Yet the trend of reducing particle density near the edge of 
deposition spot with increasing jet-to-plate distance $S/D$ 
can consistently be observed in plots (c) and (d) of Fig. 8 
for $St > St_{50}$. 

If straight nozzle is used with a jet flow at $Re = 283$, the profiles of 
$\sigma(r)$ become more or less like a Gaussian function
(similar to that in Fig. 8(d) for $S/D = 4$) 
without a clearly defined edge, in contrast to those for 
$Re = 1132$ with a flat center (cf. Fig. 3).
Relatively speaking, the tapered nozzle tends 
to deposit more particles toward the deposition spot edge 
than the straight nozzle. Reducing the value of $Re$ 
generally leads to less particle deposition toward the deposition spot edge,
i.e., the profiles of particle density $\sigma(r)$ 
exhibit more gradual slopes of decline with $r$ for smaller $Re$.

\subsection{Deposition efficiency $\eta$}

The study of particle deposition cannot be completed without 
an examination of the deposition efficiency. 
As given in Tables 2 and 3 for $Re = 1132$, 
the values of $\eta$ at a given $\sqrt{St}$ are fairly insensitive 
to the variation of the (normalized) jet-to-plate distance $S/D$,
except for straight nozzzle $\eta_{0.5}$ (with $S/D = 0.5$)
becomes noticeably lower
for $\sqrt{St} < 0.4$ and for tapered nozzle $\eta_4$ (with $S/D = 4$)
noticeably lower within a narrow interval around $\sqrt{St} = 0.47$,
e.g., from $0.42$ to $0.52$.
It seems the straight nozzle configuration is not very sensitive to 
the jet-to-plate distance variation for $S/D > 1$,
whereas the tapered nozzle configuration insensitive to  
$S/D$ variation for $S/D < 2$.
In other words,
the straight nozzles become sensitive to $S/D$ variations when $S/D < 1$ 
whereas the tapered nozzles sensitive when $S/D > 2$.

\begin{figure}[t!] \label{etavsd}
\includegraphics[scale=0.48]{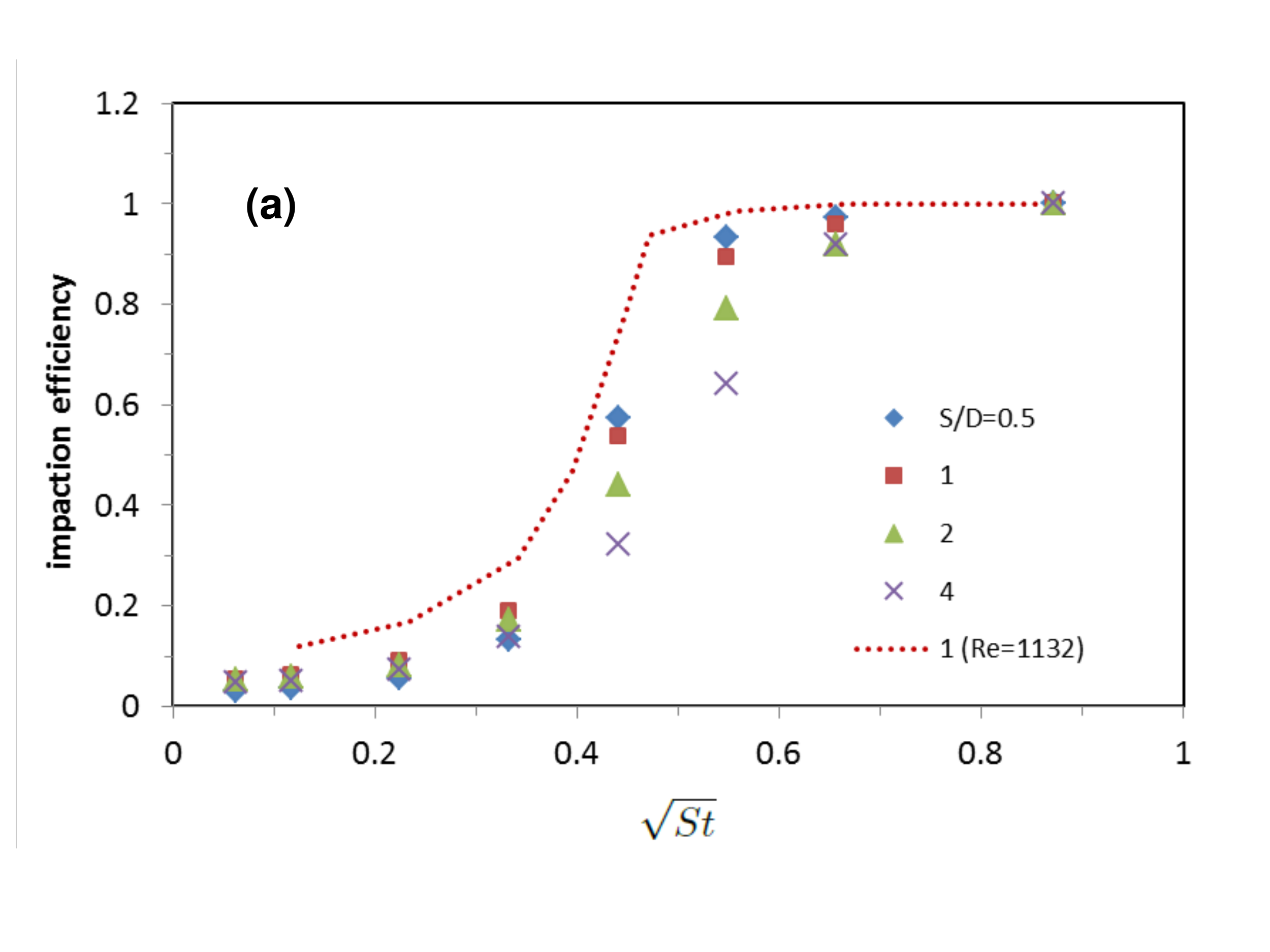}
\includegraphics[scale=0.48]{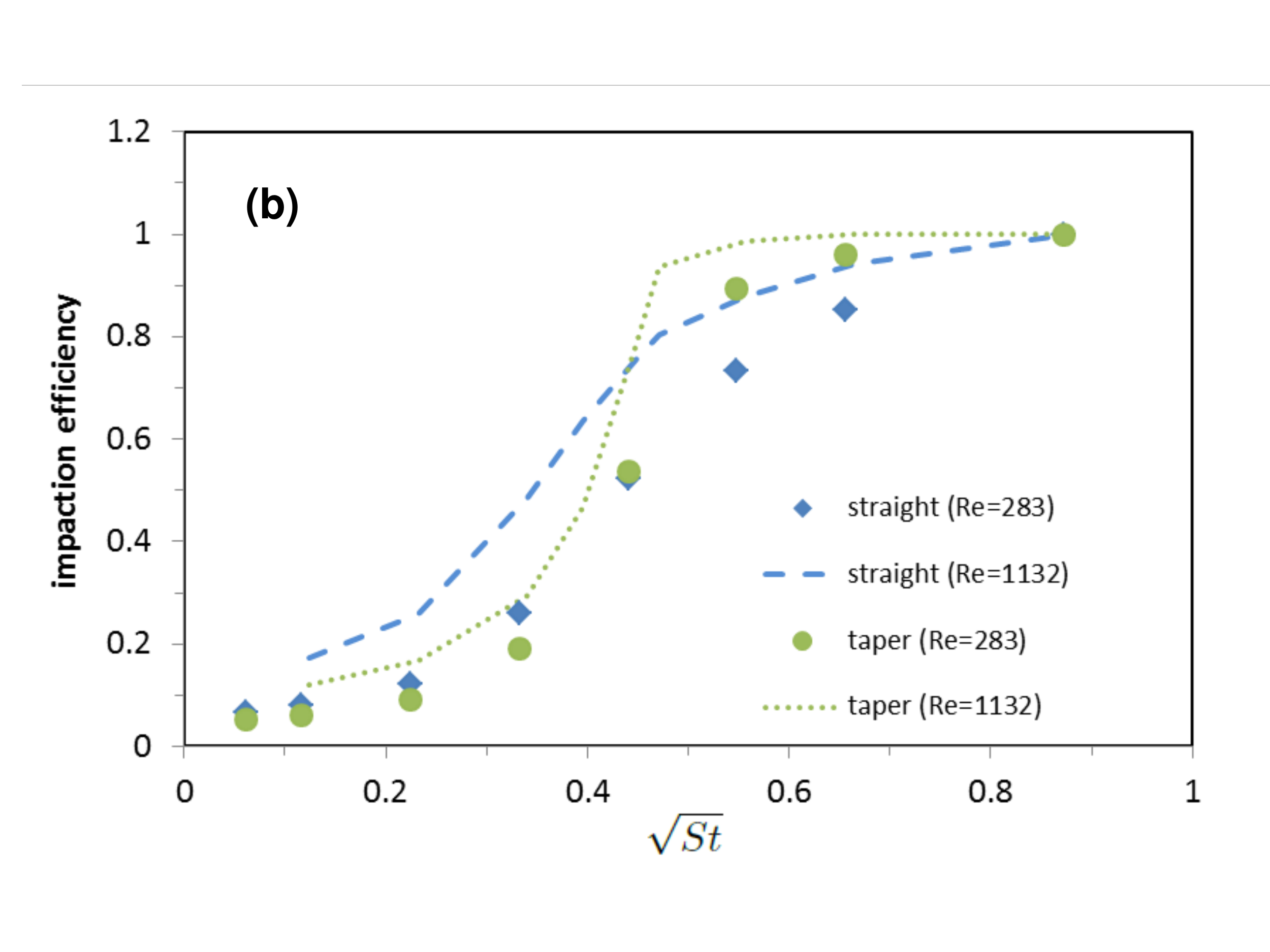}
\caption{Impaction efficiency $\eta$ versus $\sqrt{St}$. 
Plot (a) is for tapered nozzle with $\phi = 15^o$ at $Re = 283$
with $S/D = 0.5$, $1$, $2$, $4$, and at $Re = 1132$ with $S/D = 1$ as
a reference; (b) for comparison of cases with straight nozzle 
($\phi = 0^o$) and tapered nozzle ($\phi = 15^o$) at 
$Re = 283$ and $1132$ for $S/D = 1$.} 
\end{figure}

However, at $Re = 283$ the value of $\eta$ at a given $\sqrt{St}$ 
seems to change substantially with
the variation of $S/D$, especially between $\sqrt{St} = 0.4$ 
and $0.6$, as shown in
Fig. 9(a). 
A consistent trend can be seen of 
noticeably lower $\eta_4$ for relative 
larger $\sqrt{St}$ while having lower $\eta_{0.5}$
for smaller and diminishing $\sqrt{St}$, as also apparent in Tables 2---4.
The general trend of lower deposition efficiency $\eta$ 
with smaller $Re$ at a given $\sqrt{St}$ is shown in Fig. 9(b),
for both straight nozzles and tapered nozzles with the case of $S/D = 1$.
It also illustrates the fact that tapered nozzles 
yield relatively sharper cut in terms of $\eta$ versus $\sqrt{St}$,
i.e., with lower $\eta$ for $\sqrt{St} < \sqrt{St_{50}}$ 
and higher $\eta$ for larger $\sqrt{St}$, 
than the straight nozzles.

For completeness, values of deposition efficiency $\eta$ 
with straight nozzle ($\phi = 0$) at $Re = 283$ for
$S/D = 0.5$, $1$, $2$, and $4$ are computed and shown in Table 5.
Comparing with those in Table 4 for tapered nozzle,
the range of variation in $\eta$ with $S/D$ 
in Table 5 is generally smaller
with straight nozzle;  
In other words, deposition efficiency of 
straight nozzles are less sensitive to the 
jet-to-plate distance $S/D$ variations than 
that of tapered nozzles.
This could be a consequence of the lack of aerodynamic focusing
with the straight nozzle, which tends to form 
more collimated particle stream.

\begin{table}
\caption{As in Table 4  
but for taper half angle $\phi = 0$ (straight nozzle).
The values of $\sqrt{St_{50}}$ 
corresponding to $\eta = 0.5$ are $0.425$, $0.431$, $0.459$, and $0.505$
for $S/D = 0.5$, $1$, $2$, and $4$, respectrively.}
\begin{center}
\begin{tabular*}{0.75\textwidth}{@{\extracolsep{\fill}} lccccc}
\hline
\hline
\\ %\vspace{ 1 mm }
  $d_p$ ($\mu$m) & $\sqrt{St}$ & $\eta_{0.5}$ & $\eta_1$ & $\eta_2$ & $\eta_4$ \\
\hline
\\
  0.5 & $0.0622$ & $0.046$ & $0.065$ & $0.065$ & $0.054$ \\
  1 & $0.1164$ & $0.052$ & $0.079$ & $0.072$ & $0.058$ \\
  2 & $0.2245$ & $0.091$ & $0.121$ & $0.112$ & $0.094$ \\
  3 & $0.3324$ & $0.232$ & $0.259$ & $0.220$ & $0.184$ \\
  4 & $0.4404$ & $0.542$ & $0.523$ & $0.466$ & $0.363$ \\
  5 & $0.5483$ & $0.781$ & $0.734$ & $0.667$ & $0.593$ \\ 
  6 & $0.6562$ & $0.878$ & $0.853$ & $0.829$ & $0.804$ \\ 
  8 & $0.8702$ & $0.942$ & $1.000$ & $1.000$ & $1.000$ \\ 
\hline
\hline
\end{tabular*}
\end{center}
\end{table}

With regard to the small particle contamination, 
it is reduced with decreasing $Re$ as shown in Fig. 9 as well as Tables 2---4.
Yet still, the value of $\eta$ does not seem to become zero 
with diminishing $\sqrt{St}$,
for the same reason discussed in the end of section 3.1. 
At $Re = 283$, about $5\%$ of small particles with 
even $\sqrt{St} \sim 0.05$ 
seem to tenaciously impact the plate   
without being deflected by the 
radially diverging ``wall jet'' flow along the plate surface.
Apparently with optimizing the device design,
the small particle contamination may be reduced somewhat,
but cannot be eliminated completely with inertial impactors.

\section{Conclusions}
Investigation of particle deposition pattern on 
the impaction plate of an inertial impactor is carried out
numerically using a Lagrangiann solver implemented within the framework of the
OpenFOAM$^{\circledR}$
CFD package.
Various effects of the inertial impactor configuration, such as 
the jet-to-plate distance, taper angle of the nozzle channel, etc.,
are examined with discussion of physical implications.

At $Re = 1132$ (with $Re$ denoting the jet Reynolds number),
the particle deposition patterns corresponding to 
different values of $\sqrt{St}$ 
(with $St$ denoting the particle Stokes number) 
with the straight nozzle (for the taper half angle $\phi = 0$)
are mostly as expected 
with a
generally flat center and 
quick decline of the particle density 
toward the edge of the deposition spot.
But with a tapered nozzle (for $\phi = 15^o$), the deposited particles form
a high density ring near the edge of the deposition spot
especially for $St > St_{50}$, probably due to the relatively strong 
aerodynamic focusing effect on particles away from the axis.
The tapered nozzle tends to deposit particles with larger circular
patterns (with larger spot radius) 
than the straight nozzle for the same values of $St$, 
and its deposition spot radius is more sensitive to the value of $St$ 
(exppected as a result of the dependence of aerodynamic focusing effect
upon particle inertia).
A general trend of reduced value of 
the particle density peak near deposition spot edge 
is shown with increasing the jet-to-plate distance $S/D$.

With $Re$ being reduced to $283$ (as $300$ ccm flow through 
a $D = 1.5$ mm nozzle), 
particles deposited with
the same tapered nozzle ($\phi = 15^o$) 
do not seem to form the high density peak toward deposition spot edge.
As with the straight nozzle at $Re = 283$, 
reducing $Re$ tends to reduce the particle density around the 
deposition spot edge more than that closer to the axis.
The same trend also applies to the effect of increasing 
the jet-to-plate distance $S/D$ in general.

A close examination of the particle deposition efficiency $\eta$ 
shows the fact that a small amount of very fine particles with 
extremely small values of $St$ always impact the center of plate.  
Thus, the value of $\eta$ does not approach zero with 
a substantial (orders of magnitude) reduction of the value of $St$.
It should not be difficult to understand that 
particles along the axis ($r \sim 0$) will not be easily deflected by the 
sharply bending streamlines, with diminishing magnitude of 
the radial velocity component around the center of the stagnation zone.
Such a ``small particle contamination'', 
which typically amounts to $\sim 10\%$ small particles 
with $\sqrt{St} < 0.1$ at $Re \sim 1000$ and 
$\sim 5\%$ at $Re \sim 300$, 
may not be negligible in data analysis 
with inertial impactor measurement.

\section*{Acknowledgments}
The author would like to thank John Lees for 
support and guidance, and John Hamre,
Dr. Kurt Christenson, Dr. Mike Renn, as well as 
many other Optomec colleagues, for 
helpful technical discussions.

\vspace{4 mm}

%\label{lastpage}

\end{document}